%% file: main.tex
\documentclass[10pt,a4paper,twocolumn]{article}
\usepackage{mathptmx} 
\usepackage{graphicx} 
\usepackage{amstext} 
\usepackage{amsmath}
\usepackage{amssymb}
\usepackage{booktabs}
\usepackage{multicol}
\usepackage{subfigure}
\include{orcid}
\usepackage[top=25mm,bottom=37mm,left=20mm,right=20mm,columnsep=10mm]{geometry} 
\usepackage{color} 
\definecolor{myblu}{rgb}{0.1,0.1,0.5}
\usepackage{hyperref} 
\hypersetup{colorlinks=true,urlcolor=blue,linkcolor=black,citecolor=black} 
\usepackage{sectsty,textcase} 
\sectionfont{\large\MakeTextUppercase} 
\usepackage{secdot} 
\usepackage{amsmath,bm}

\begin{document}
\title{\vspace{-18mm}
\begin{minipage}{\linewidth}
\hspace{5mm}\raisebox{-50pt}{\includegraphics[width=.23\textwidth]{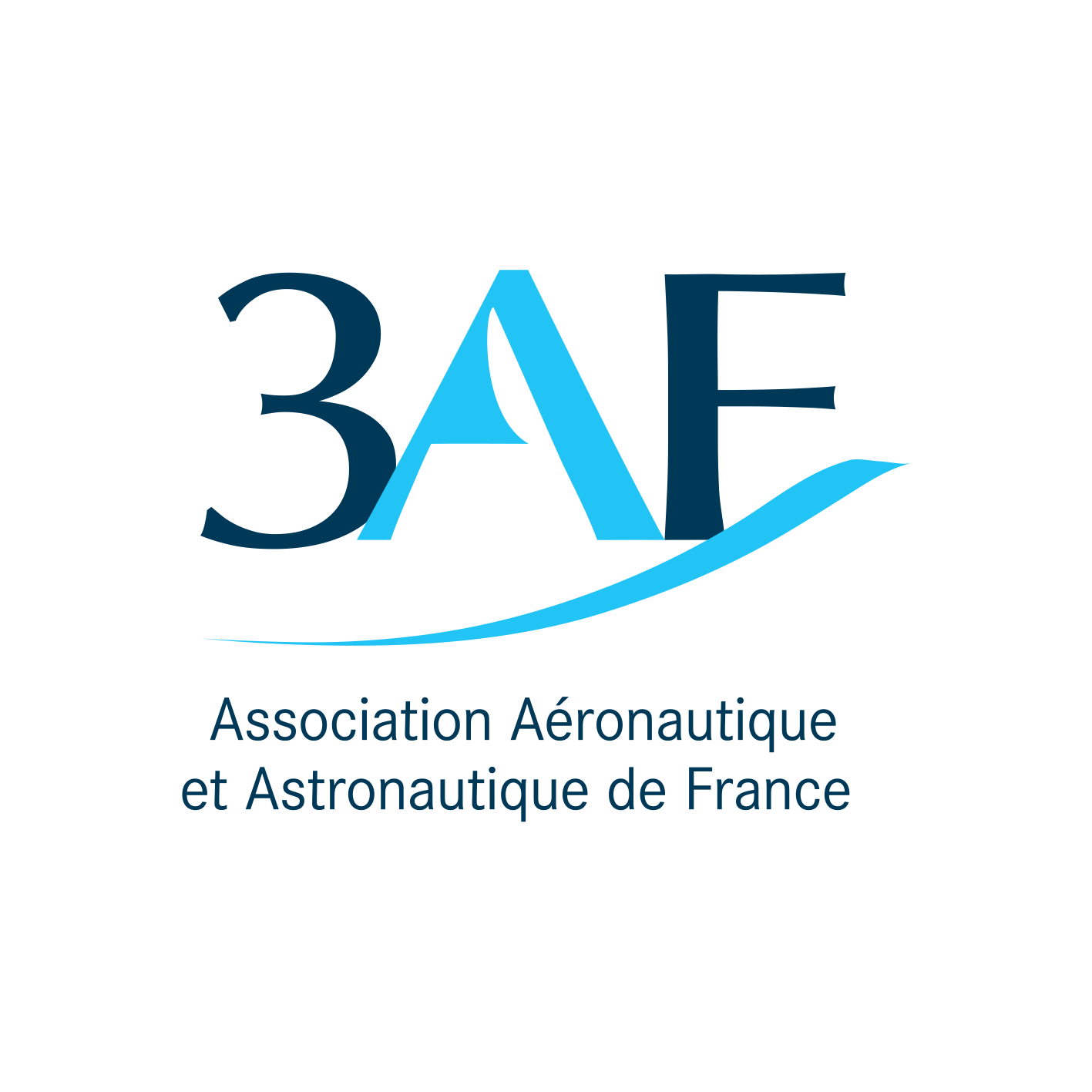}}\hspace{4mm}
\textcolor{myblu}{\textbf{\textit{\normalsize\begin{tabular}{l}
57$^\text{th}$ 3AF International Conference\\ on Applied Aerodynamics\\ 29 --- 31 March 2023, Bordeaux -- France
\end{tabular}}}}
\hspace{16mm}\textbf{\normalsize AERO2023-ID-NAME}
\end{minipage}\\\vspace{10mm}
\textbf{\Large AEROTHERMODYNAMIC SIMULATORS FOR ROCKET DESIGN \\ USING NEURAL FIELDS}}
\author{\textbf{\normalsize Haitz Sáez de Ocáriz Borde$^\text{(1)}$, Pietro Innocenzi$^\text{(2)}$ and Flavio Savarino \orcidicon{0000-0002-2576-0685}$^\text{(3)}$}
\\{\normalsize\itshape
$^\text{(1)}$Oxford Robotics Institute, University of Oxford, United Kingdom, haitz@oxfordrobotics.institute}
\\{\normalsize\itshape
$^\text{(2)}$Dep. of Mechanical Engineering, Imperial College London, United Kingdom, pietro.innocenzi20@imperial.ac.uk}
\\{\normalsize\itshape
$^\text{(3)}$Dep. of Aeronautics, Imperial College London, United Kingdom, flavio.savarino17@imperial.ac.uk}}
\date{}

\maketitle
\begin{abstract}
The typical size of computational meshes needed for realistic geometries and high-speed flow conditions makes Computational Fluid Dynamics (CFD) impractical for full-mission performance prediction and control. Reduced-Order Models (ROMs) in low-speed aerodynamics have come a long way in terms of reconstructing coherent flow patterns, thus enabling aerodynamic performance prediction. While many examples of ROMs exist for low-speed aerodynamics, there is no such broad literature for high-speed flows. We propose to use physics-enhanced neural fields for prediction of the steady, supersonic flow around a rocket for resolving: the bow shock profile, the boundary layer gradients and the wake over a range of incidences. This approach can lead to the construction of a computationally inexpensive, continuous aerothermodynamic model of a rocket at realistic flight conditions with applications to aerodynamic design and mission analysis and control. The use of neural fields allows to quickly and efficiently sweep the Angle of Attack (AoA) in a continuous manner, as compared to traditional CFD which requires running simulations for each discrete incidence value.
\end{abstract}

\section{Introduction}\label{sec: Introduction}
In aerodynamics, parametric Reduced-Order Models~(ROMs) are used to predict the flow field distribution based on high-fidelity Computational Fluid Dynamics~(CFD), but with faster evaluation time and requiring less computational memory as compared to the original CFD model~\cite{Ripepi2018ReducedorderMF}. In this work we use neural fields~\cite{Xie2021NeuralFI,Mildenhall2020NeRFRS} based ROMs to obtain aerothermodynamics simulators trained based on CFD data. These are able to predict the pressure, velocity, and temperature distributions around a rocket at arbitrary angles of attack~(AoAs). We also incorporate well-known aerothermodynamics physics relations such as the no-slip boundary condition and the temperature-Mach number relation for adiabatic flow into the models, and evaluate their performance and simulation capabilities. 

The use of neural fields in aerodynamics provides several advantages over traditional CFD simulations. One of the main benefits is the ability to interpolate quickly and efficiently in terms of the AoA. This process can be performed in a matter of seconds, which is significantly faster than the time required to run a CFD simulation. This speed is especially beneficial for design iterations and optimization, where quick and accurate results are crucial. Another advantage of neural fields is the generation of continuous interpolations. Unlike traditional methods that only provide discrete solutions, neural fields can produce smooth and continuous solutions. This is beneficial when analyzing complex aerodynamic systems, where the flow field can exhibit complex behavior that changes gradually over the domain.

In this work, we use the AoA as the sweeping parameter for our neural field model. However, it is possible to interpolate in terms of other parameters such as rocket geometry. This would enable the rapid exploration of the design space and could drastically speed up the aerothermodynamic design process. By using neural fields, engineers and designers can quickly and efficiently evaluate different design options and make informed decisions about the optimal configuration of their system.

The rest of the paper is organized as follows: in Section~\ref{sec: Related Work} we discuss related work regarding ROMs and Deep Learning (DL)~\cite{LeCun2015DeepL} for aerospace engineering; in Section~\ref{sec: Background} we review relevant background on supervised machine learning and neural fields; in Section~\ref{sec: Discrete Aerothermodynamics Data Generation} we explain the aerothermodynamics data generation process, in particular we work with a 2D rocket geometry at relatively low AoAs; in Section~\ref{sec: Continuous Reduced Order Models using Physics-Informed Neural Fields} we describe the main method used to obtain ROMs for aerothermodynamics using neural fields, as well as how to incorporate physics into the models; Section \ref{sec: Results} presents the results alongside a thorough ablation study and aerothermodynamic evaluation; and lastly, in Section~\ref{sec: Conclusion} we consider the final conclusions and propose future research directions.

\section{Related Work}\label{sec: Related Work}
In aerodynamics, ROMs are simplified mathematical models that capture the essential behavior of complex aerodynamic systems while reducing the computational complexity of the analysis. These models are used to accelerate the design process, reduce the computational cost of simulations, and enable real-time control of complex systems. The basic idea behind ROMs is to identify the most important features of a system and use them to construct a simplified model that captures the dominant behavior. Reduced order models have a variety of applications in aerodynamics, including the design of aircraft, cars, and other vehicles, the optimization of control systems, and the prediction of aerodynamic loads on structures. By reducing the computational complexity of the analysis, ROMs enable engineers and scientists to explore a wider range of the design spectrum.

In the literature, both unsteady laminar~\cite{Durmaz2013,Stabile2017,Juan2019} and turbulent~\cite{Nidhan2020spodwake,Schmidt2018jet,Fukami2021,Giannopoulos2020} flow regimes have been analysed using algorithms able to extract coherent structures and patterns to inform dynamical models with predictive and control capabilities. Linear models have demonstrated good performance in nonlinear and highly-nonlinear, large-scale systems, but necessitate of real-time, partial measurements of the original system dynamics to learn the contribution of the nonlinearities~\cite{Juan2019,Papadakis2021,Loiseau2018,savarino_papadakis_2022}. On the other hand, nonlinear models~\cite{Nair_Goza_2020,Kim_kim_won_lee_2021,Rozon_Breitsamter_2021} result in more sophisticated architectures but are prone to tuning issues and loss of interpretability.

Concurrently, DL has been applied to a number of problems in aerospace engineering, such as turbulence modeling~\cite{Ling2016ReynoldsAT,Borde2021ConvolutionalNN,multitask,Sayyari2022UnsupervisedDL,Fang2019NeuralNM,Kaandorp2018MachineLF}, aerodynamic shape optimization~\cite{Li2022MachineLI,Xu2021MachineLF,Li2021OnDG,Liu2023DeeplearningbasedAS}, wall-flux-based wall models~\cite{LozanoDuran2020SelfcriticalMW}, and rocket liquid engine design~\cite{WaxeneggerWilfing2021MachineLM} to name a few. DL is able to approximate a wide range of relationships leveraging neural networks which can capture complicated nonlinear relationships based on data. In essence, artificial neural networks can be used to learn an aerothermodynamics model which is captured within the parametrization of the networks. 

As stated earlier, our models are based on a database of CFD simulations that analyze the rocket geometry at different AoAs, see Section~\ref{sec: Discrete Aerothermodynamics Data Generation}. One of the main drawbacks of artificial neural networks is their black-box nature, which leads them to only match the training data without taking into account the physical principles that produced it. In this study, we incorporate known physics relationships to improve the performance of the network and direct it towards solutions that are physically plausible and meaningful. After training, the neural network can rapidly predict the flow field for any AoA. However, it should be noted that the model cannot forecast flow separation because it was not trained on high AoAs.

\section{Background}\label{sec: Background}
In this section we provide relevant background regarding artificial neural networks and supervised learning, neural fields, Fourier positional feature mappings, and multi-task learning.

\subsection{Supervised Learning and Artificial Neural Networks}\label{subsec: Supervised Learning and Artificial Neural Networks}
Supervised machine learning is a function estimation problem~\cite{LeCun2015DeepL}. Models must approximate an unknown function, $f : \mathbb{R}^{a}\rightarrow\mathbb{R}^{b}$, based on observed input-output pairs (where $a$ and $b$ are the input and output dimensionalities). DL uses artificial neural networks to obtain an approximate function $f^{*} : \mathbb{R}^{a}\rightarrow\mathbb{R}^{b}$ that best describes the observed data, and ideally, generalizes well to new examples. The function $f^{*}$ is based on a composition of other nonlinear functions, i.e. the layers of the model:
\begin{equation}
   f^{*}= l^{n}\circ l^{n-1}\circ ... \circ l^{1}, 
\end{equation}
where the operator $\circ$ is used to denote function composition, such that two functions say $f_1$ and $f_2$ generate a new function $f_3$ so that $f_3(x) = f_2 \circ f_1(x)= f_2(f_1(x))$, and $l^{i}$ refers to the $i^{th}$ layer: $l^{1} : \mathbb{R}^{a}\rightarrow\mathbb{R}^{h_1}$, $l^{i} : \mathbb{R}^{h_{i-1}}\rightarrow\mathbb{R}^{h_i}\,\forall i\in [2,n-1]$, and $l^{n} : \mathbb{R}^{h_{n-1}}\rightarrow\mathbb{R}^{b}$. $h_i$ is the hidden dimension for the $i^{th}$ layer. The model parameters are optimized based on a loss function. In our case, we use the mean-squared error (MSE) loss to update the model parameters by comparing the model prediction output against the ground truth aerodynamic data obtained using CFD.

The model learns a continuous function from the discrete snapshots provided by the CFD simulations. This avoids the computational hurdle of having to recompute the flow solution for every possible AoA using CFD, which in our case is the sweeping parameter. As previously mentioned in Section~\ref{sec: Introduction}, other sweeping parameters describing the rocket geometry, Mach number, or flow properties could potentially be added in the future. The neural network function can be expressed mathematically as:
\begin{equation}
    f^{*}(x,y,\alpha)=(p,T,v_{x},v_{y}),
\end{equation}
where $p$ is the static pressure, $T$ is the static temperature, and $v_{x}$ and $v_{y}$ are the velocity components for a given coordinate $(x,y)$ in 2D and rocket AoA, $\alpha$.

\subsection{Neural Fields and Fourier Positional Feature Mappings}\label{subsec: Neural Fields and Fourier Positional Feature Mappings}
Neural fields have gained traction as a new approach to signal representation in a range of domains, including computer vision, computer graphics, and other areas~\cite{Xie2021NeuralFI}. We can define a field as a quantity which is given for all spatial or temporal coordinates, and a neural field as a field that is parameterized by a neural network that maps input coordinates to a scalar or tensor output. A Neural Radiance Field (NeRF)~\cite{Mildenhall2020NeRFRS} is a type of deep neural network architecture, and a specific class of neural field used for 3D scene reconstruction from 2D images. It uses a Multi-Layer Perceptron (MLP) to model the mapping from a set of input image features to a set of 3D radiance values. An MLP is a simple feedforward neural network composed of multiple layers of artificial neurons, where each layer is fully connected to the previous one. In our case, instead of modeling the 3D scene in terms of $(x,y,z)$ coordinates, we can think of the AoA as being the third dimension instead, and map coordinates to aerothermodynamic quantities.

Fourier features have been used in the literature to allow MLPs to learn high-frequency functions in low-dimensional problem domains such as image regression, 3D shape regression, MRI reconstruction, and inverse rendering~\cite{Tancik2020FourierFL}. Fourier positional feature mappings can be used to improve the performance of the neural field by encoding the position of a point in space as a series of sine and cosine functions with different frequencies, which are then concatenated with the input features of the network. This can help the network better capture the spatial structure of the scene, leading to more accurate and detailed reconstructions. In this work we use Gaussian mappings
\begin{equation}
    \Gamma(\mathbf{v}) = [\cos(2\pi\mathbf{B}\mathbf{v}),\sin(2\pi\mathbf{B}\mathbf{v})]^{T},
\end{equation}
where $\mathbf{v}$ represents the coordinates $x$ and $y$, and $\mathbf{B}\in\mathbb{R}^{m\times 2}$ is a random projection matrix sampled from $\mathcal{N}(0,\sigma^{2})$ in line with the literature~\cite{Tancik2020FourierFL}. We experimentally adjust $m$ and $\sigma^{2}$. This effectively means that the network relationship can be rewritten as
\begin{equation}
    f^{*}(\Gamma(\mathbf{v}),\alpha)=(p,T,v_{x},v_{y}).
\end{equation}

\subsection{Multi-Task Learning}\label{subsec: Multi-Task Learning}
Multitask learning is a machine learning technique where a single model is trained to perform multiple related tasks simultaneously. The tasks share some common features or underlying structure, and the model learns to leverage this shared knowledge to improve its performance on all tasks~\cite{Zhang2017ASO}. 

One of the benefits of multitask learning is that it can lead to easy convergence during training. This is because the shared knowledge between tasks can act as a regularizer, which helps to prevent overfitting and improves the generalization performance of the model. By jointly optimizing multiple tasks, the model can learn a more robust and generalizable representation of the input data. This is achieved by the model learning to extract features that are relevant to all tasks, rather than learning task-specific features that may not generalize well to other tasks. As a result, the model can learn more efficiently and converge faster than training separate models for each task. 

In addition to the regularization effect, multitask learning can also benefit from the transfer learning effect, where the model can transfer knowledge learned from one task to another. This can be especially beneficial when training data is limited or expensive to acquire, as the model can leverage the knowledge gained from one task to improve its performance on another related task. 

Multi-task learning has been applied to various machine learning tasks such as Computer Vision~\cite{Kendall2017MultitaskLU}, Natural Language Processing~\cite{Zhang2022ASO}, and Speech Recognition~\cite{Mo2022MultiTaskLI}, and has shown promising results in improving performance and reducing training time and computational resources~\cite{Ruder2017AnOO}. In particular, in this work we will implement hard parameter sharing multi-task learning in line with previous work combining machine learning and turbulence modeling~\cite{multitask}. This consists in sharing initial layers upstream of the model, for all tasks (pressure, temperature, and velocity distribution estimations), and using that shared information to feed a few specialized layers for each of the predictions. Unlike a standard artificial neural network, employing a multi-task learning architecture allows to optimize the parameters of the last layers of the network based on independent losses: $L_{p},L_{T},L_{v_x},L_{v_y}$, while still updating most of the network parameters upstream of the model based on a compound loss 
\begin{equation}
    L_{total}=L_{p}+L_{T}+L_{v_x}+L_{v_y}.
\end{equation}

In other words, some of the parameters for the last layers of the network are only updated based on single property prediction. This provides extra flexibility to the network during optimization. 

Mathematically we can express the learned function, $f^{*}(\Gamma(\mathbf{v}),\alpha)=(p,T,v_{x},v_{y})$ as
\begin{equation}
    p = f_p\circ g(\Gamma(\mathbf{v}),\alpha),
\end{equation}
\begin{equation}
    T = f_T\circ g(\Gamma(\mathbf{v}),\alpha),
\end{equation}
\begin{equation}
    v_x = f_{v_x}\circ g(\Gamma(\mathbf{v}),\alpha),
\end{equation}
\begin{equation}
    v_y = f_{v_y}\circ g(\Gamma(\mathbf{v}),\alpha).
\end{equation}

In the expression above we have different functions for each of the components: $f_p,f_T,f_{v_x},f_{v_y}$, but all take as input with the same representation produced by the function $h_{latent}=g(\Gamma(\mathbf{v}),\alpha)$. This effectively means that the weights used to parameterize the function $g$ affect the predictions for all components, whereas those in each of the other functions $f_p,f_T,f_{v_x}$, and $f_{v_y}$ are independent. This has important implications during backpropagation. Let us define $\boldsymbol{\Theta}_g$ as the set of weights used to parameterize the function $g$: for the partial derivative 
\begin{equation}
    \frac{\partial L_{total}}{\partial \boldsymbol{\Theta}_g} = \frac{\partial L_{p}}{\partial \boldsymbol{\Theta}_g}+\frac{\partial L_{T}}{\partial \boldsymbol{\Theta}_g}+\frac{\partial L_{v_x}}{\partial \boldsymbol{\Theta}_g}+\frac{\partial L_{v_y}}{\partial \boldsymbol{\Theta}_g},
\end{equation}
when expanded, all its summation components
\begin{equation}
    \frac{\partial L_{p}}{\partial \boldsymbol{\Theta}_g}=\frac{\partial L_{p}}{\partial f_p}\frac{\partial f_p}{\partial g}\frac{\partial g}{\partial \boldsymbol{\Theta}_g},
\end{equation}
\begin{equation}
    \frac{\partial L_{T}}{\partial \boldsymbol{\Theta}_g}=\frac{\partial L_{T}}{\partial f_T}\frac{\partial f_T}{\partial g}\frac{\partial g}{\partial \boldsymbol{\Theta}_g},
\end{equation}
\begin{equation}
    \frac{\partial L_{v_x}}{\partial \boldsymbol{\Theta}_g}=\frac{\partial L_{v_x}}{\partial f_{v_x}}\frac{\partial f_{v_x}}{\partial g}\frac{\partial g}{\partial \boldsymbol{\Theta}_g},
\end{equation}
\begin{equation}
    \frac{\partial L_{v_y}}{\partial \boldsymbol{\Theta}_g}=\frac{\partial L_{v_y}}{\partial f_{v_y}}\frac{\partial f_{v_y}}{\partial g}\frac{\partial g}{\partial \boldsymbol{\Theta}_g},
\end{equation}
are non-zero. Hence to update the weights we would use the following expression:
\begin{multline}
    \boldsymbol{\Theta}^{t}_g = \boldsymbol{\Theta}^{t-1}_g -\eta\frac{\partial L_{total}}{\partial \boldsymbol{\Theta}^{t-1}_g} \\ = \boldsymbol{\Theta}^{t-1}_g -\eta\left(\frac{\partial L_{p}}{\partial \boldsymbol{\Theta}^{t-1}_g}+\frac{\partial L_{T}}{\partial \boldsymbol{\Theta}^{t-1}_g}+\frac{\partial L_{v_x}}{\partial \boldsymbol{\Theta}^{t-1}_g}+\frac{\partial L_{v_y}}{\partial \boldsymbol{\Theta}^{t-1}_g}\right),
\end{multline}
where we use $t$ to refer to the optimization time step during training and $\eta$ for the learning rate. In the expression above, $\boldsymbol{\Theta}^{t}_g$ would correspond to the set of weights of the function $g$ in time step $t$, which are update based on the weights in the previous iteration $\boldsymbol{\Theta}^{t-1}_g$. Note that in practice we use the Adam optimizer~\cite{Kingma2014AdamAM}, but for simplicity we display the standard derivation used for stochastic gradient descent. However, in the case of the rest of the functions, denoting $\boldsymbol{\Theta}_p,\boldsymbol{\Theta}_T,\boldsymbol{\Theta}_{v_x},\boldsymbol{\Theta}_{v_y}$ as the sets of weights for each function $f_p,f_T,f_{v_x},f_{v_y}$, respectively, the update functions would become
\begin{multline}
    \boldsymbol{\Theta}^{t}_i = \boldsymbol{\Theta}^{t-1}_i -\eta\frac{\partial L_{total}}{\partial \boldsymbol{\Theta}^{t-1}_i} \\ = \boldsymbol{\Theta}^{t-1}_i -\eta\frac{\partial L_{i}}{\partial \boldsymbol{\Theta}^{t-1}_i};\,i\in{[p,T,v_x,v_y]}.
\end{multline}
Note that in this case each set of weights its only updated based on a component of the total loss, $L_{total}$.

\section{Discrete Aerothermodynamics Data Generation}\label{sec: Discrete Aerothermodynamics Data Generation}
The commercial CFD solver STAR-CCM+ is used to solve the steady 2D laminar Navier-Stokes equations around a simplified rocket geometry at Mach number $M_{\infty}=2.5$, and we use the MUSCL central differencing scheme with $3^{rd}$ order accuracy in space for flux discretization. The inviscid fluxes are evaluated via the AUSM+ flux-vector splitting scheme \cite{ausm}. The working fluid is air, which is modelled as an ideal gas ($\gamma=c_p/c_V=1.4$). The inlet flow conditions are $M_{\infty}=2.5$ and flow properties at atmospheric altitude $h$~=~25\,km. Freestream static values of temperature, $T_{\infty}$, and pressure, $p_{\infty}$, are obtained from the International Standard Atmosphere (ISA) model, leading to the rocket diameter-based Reynolds number $Re_{D}=2.76\times10^5$. This value justifies the choice of simulations in the laminar regime \cite{tran}.  The wall is modelled as isothermal with $T_{wall} = 300 \mathrm{K}$. Data is obtained for the geometry at these conditions and several AoAs in the range $0^{\circ}\leq\alpha\leq12^{\circ}$.

The computational domain is 2D-rectangular of length 20$L$ and height 100$D$, where $L$ = 120\,mm and $D$ = 10\,mm are the rocket length and diameter, respectively. The nosecone profile is elliptical with semi-major axis $r_{1}$ = 20\,mm and semi-minor axis $r_{2}$ = 5\,mm. The half ellipse is attached to the main body tube of the rocket, which appears as a rectangle of length-to-height ratio 10.

\begin{figure}[hbt!]
\centering
\includegraphics[width=0.45\textwidth]{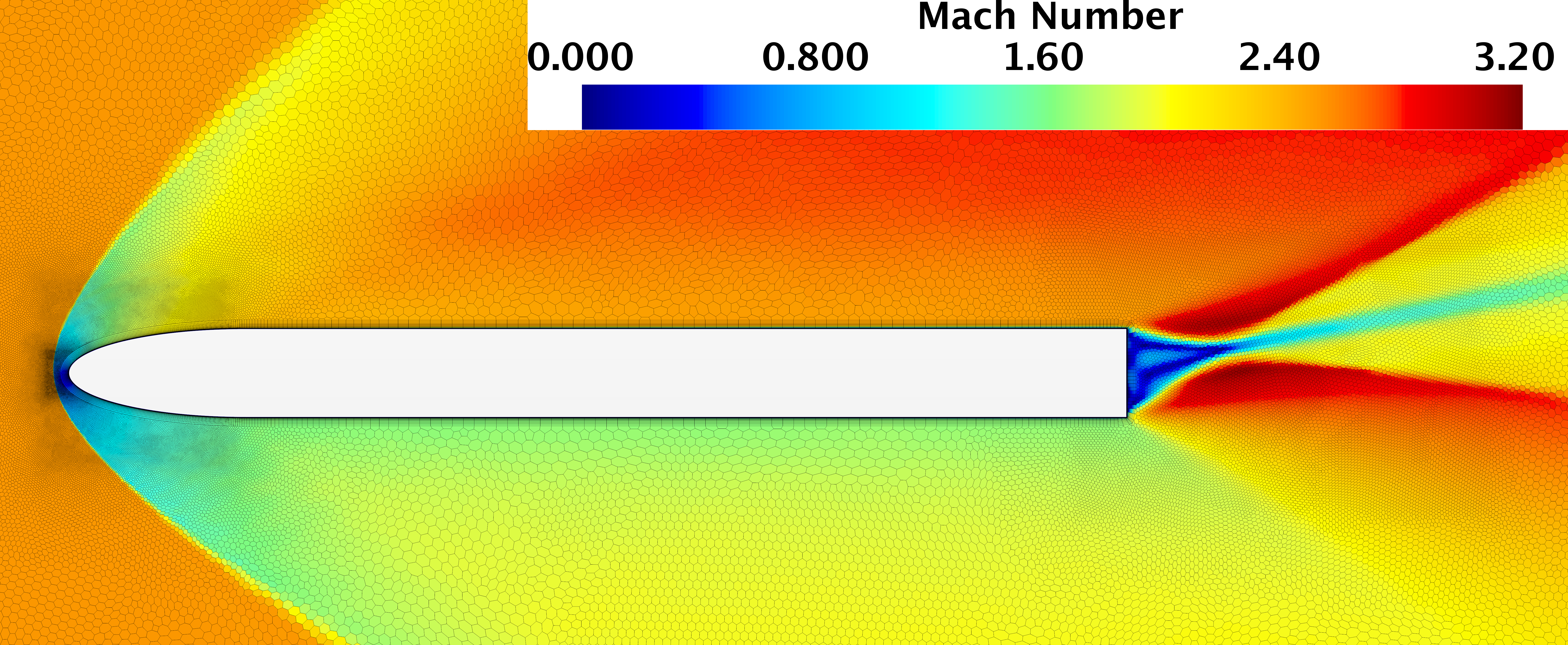}
\caption{Converged 2D mesh of the simplified rocket geometry, with Mach number $M = 2.5$, and AoA equal to $\alpha = 10^{\circ}$.}
\label{fig:mesh}
\end{figure}

The mesh is generated within STAR-CCM+ and consists of prism layer cells near viscous boundaries surrounded by polyhedral mesh elsewhere. A mesh convergence study with respect to residuals and aerodynamic coefficients of lift, drag, pitch moment as well as with respect to the thermal boundary layer values at multiple locations was first carried out, leading to a final converged mesh of around 250 x 10\textsuperscript{3} points. The final mesh is shown in Fig.~\ref{fig:mesh}, superimposed on the computed Mach number field for one of the simulation cases.

\begin{figure}[hbt!]
\centering
\includegraphics[width=0.45\textwidth]{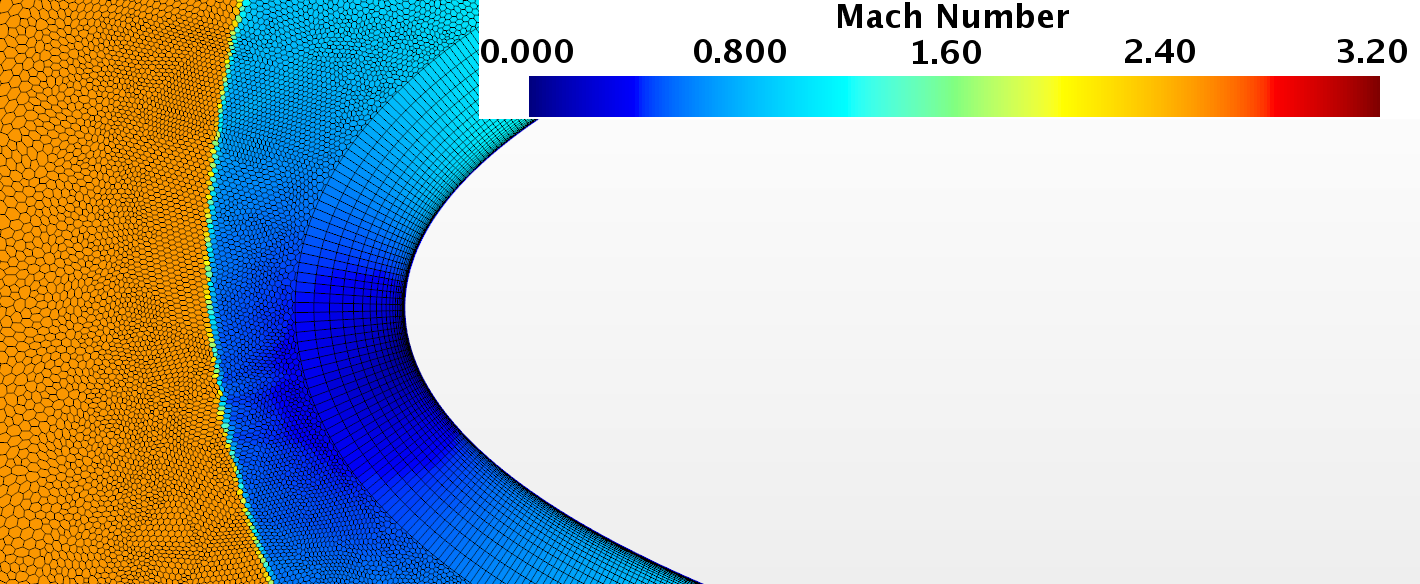}
\caption{Volumetric mesh refinement around the nosecone. Simulation with Mach number $M = 2.5$, and AoA equal to $\alpha = 10^{\circ}$.}
\label{fig:nose}
\end{figure}

\begin{figure}[hbt!]
\centering
\includegraphics[width=0.45\textwidth]{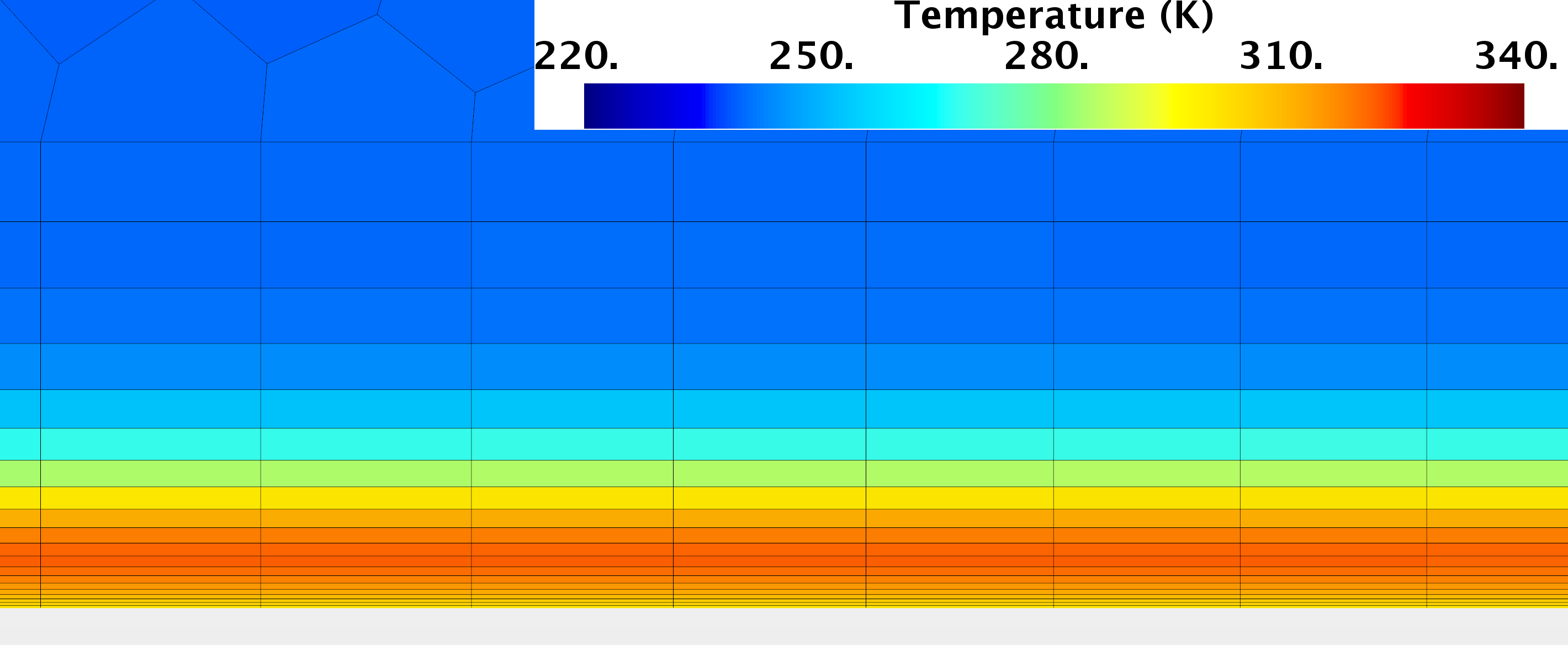}
\caption{Prism layer mesh and thermal boundary layer. Simulation with Mach number $M = 2.5$, and AoA equal to $\alpha = 10^{\circ}$.}
\label{fig:bl1}
\end{figure}

The bow shock in front of the nosecone is refined with a volumetric refinement approach to obtain accurate pressure values at the stagnation region, and the prism layer mesh is refined to accurately compute the gradients at the wall, see Fig.~\ref{fig:nose} and \ref{fig:bl1}. The first cell Reynolds number, defined as:
\begin{equation}
    Re_{c} = \frac{\rho_{c} a_{c} \Delta y}{\mu_{c}},
    \label{reCell}
\end{equation}
where $\rho_{c}$, $a_{c}$ and $\mu_{c}$ are the density, speed of sound and viscosity of the fluid at the cell and $\Delta y$ is the height of the first prism layer cell from the wall, never increases beyond 12 throughout the geometry. Such value proved to be enough to allow mesh independence of the thermal boundary layer results \cite{papadopoulos}. 

\section{Continuous Reduced Order Models using Physics-Informed Neural Fields}\label{sec: Continuous Reduced Order Models using Physics-Informed Neural Fields}
In this section we present several techniques to incorporate known physical flow properties into the network architecture. This will help neural networks, which in principle are black-boxes with no physics knowledge, converge faster and towards more accurate solutions, that are also easier to interpret.

\subsection{No-Slip Boundary Condition Enforcement} \label{subsec: No-Slip Boundary Condition Enforcement}
The no-slip boundary condition is a fundamental concept in fluid dynamics that describes the behavior of a fluid in contact with a solid surface. It states that at the interface between a fluid and a solid surface, the fluid velocity relative to the surface must be zero~\cite{Anderson1984FundamentalsOA,Tropea2007MicrofluidicsT}.

We enforce the no-slip boundary condition for the velocities $v_x$ and $v_y$. That is, the velocity on the rocket wall must be zero. To do so, we use a positional one-hot encoding $w_i$, where $w_i=1$ at the wall and $w_i=0$ elsewhere. We multiply the predicted velocities as follows: 
\begin{equation}
    v_x=(1-\lambda(\beta_1)w_i)\tilde{v}_x,
    \label{vx_bc}
\end{equation}
\begin{equation}
v_y=(1-\lambda(\beta_2)w_i)\tilde{v}_y,
\label{vy_bc}
\end{equation}
where $\lambda(\beta) = \frac{1}{1+e^{-\beta}}$ is the sigmoid activation function, $\beta_1$ and $\beta_2$ are learnable hyperparameters updated based on the loss function during training, and $\tilde{(\cdot)}$ denotes the original network prediction. The sigmoid function is bounded $0<\lambda(\beta)<1$, and allows the model to learn the importance of the boundary condition enforcement during training. Otherwise, we observe that imposing hard no-slip boundary condition enforcement, such as $v_x=(1-w_i)\tilde{v}_x$ and $v_y=(1-w_i)\tilde{v}_y$, leads to artefacts during training. For training we initialize $\beta_1=-3$ and $\beta_2 = -2$. This is in line with similar techniques discussed in the literature~\cite{Borde2021ConvolutionalNN,multitask}.

Additionally, since the CFD simulations assumed an isothermal wall with $T_{wall} = 300K$, we also enforce the value of temperature at the wall:
\begin{equation}
    T = \tilde{T}(1-\lambda(\beta_3)w_i)+\lambda(\beta_3)T_{wall}w_i,
    \label{bc_temp}
\end{equation}
where $\beta_3$ is another learnable hyperparameter. We initialize $\beta_3 = -2$ at the beginning of training. We can consider this to be a $\textit{soft}$ boundary condition enforcement.

\subsection{Inductive Bias for Temperature Prediction}\label{subsec: Inductive Bias for Temperature Prediction}
Next, we outline the method used to incorporate inductive biases for the temperature prediction away from the wall into the networks.

\subsubsection{Temperature-Mach number relation Justification}\label{subsubsec: Temperature-Mach number relation Justification}
\begin{figure}
    \centering
    \includegraphics[width=0.45\textwidth]{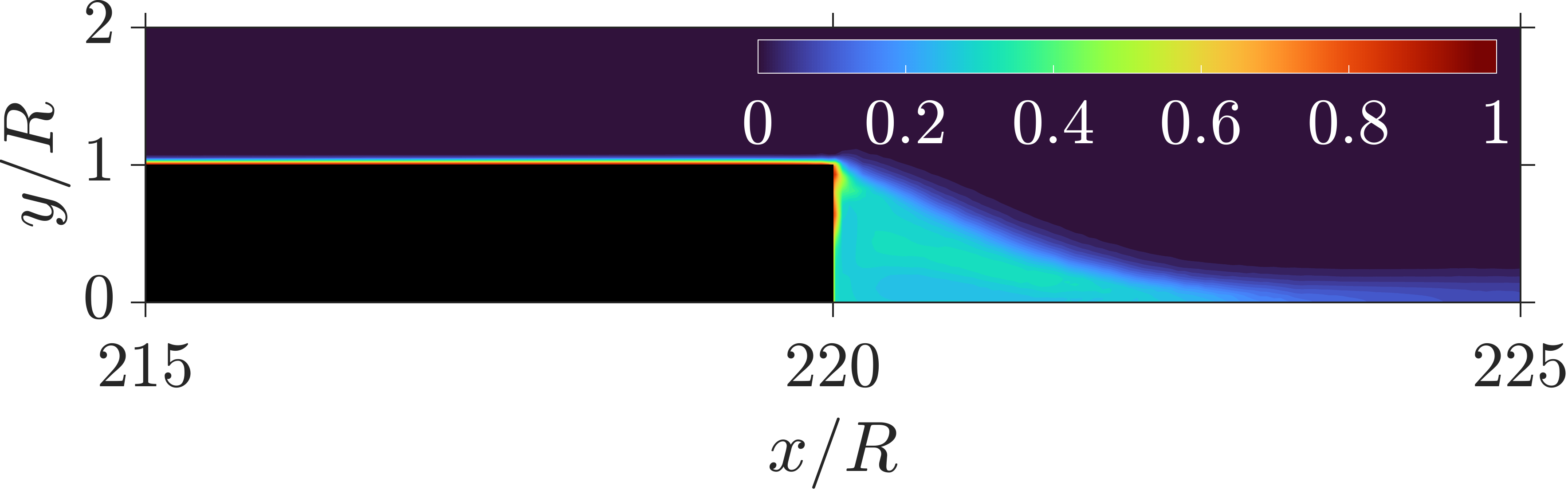}
    \caption{Contours of absolute error between the true, $\hat{T}$, and adiabatic temperature approximation, $T_a$, normalised by the free-stream temperature, $T_\infty$. The black-coloured shape is the trailing section of the body. Spatial coordinates are normalised by the rocket radius, $R$.}
    \label{fig:error contours}
\end{figure}

Although the simulated flow field is not adiabatic, we consider the temperature-Mach number relation for adiabatic, compressible flows at stagnation temperature condition $T_0$,
\begin{equation}
    T_a = \frac{T_0}{\left(1 + \frac{\gamma-1}{2}M^2\right)},
    \label{eq:adiabatic T}
\end{equation}
as a useful physical relation to be applied only in suitable regions of validity, that is, outside viscous boundary layers and the wake. The approximate temperature field, $T_a$, can be computed directly from Eq.~\ref{eq:adiabatic T} by knowing the Mach number, $M$, and the stagnation temperature, $T_0$. The former is given by the CFD data, while the latter by the imposed boundary conditions. We visualise the normalised error between the true temperature from CFD, $\hat{T}$, and the adiabatic approximation, $T_a$, in Figure~\ref{fig:error contours} for the case $\alpha=0^{\circ}$, showing only the upper half section of the geometry for symmetry reasons. As expected, the error is significant in the near-wall boundary layer region and modest inside the wake.

\begin{figure}
    \centering
    \includegraphics[width=0.45\textwidth]{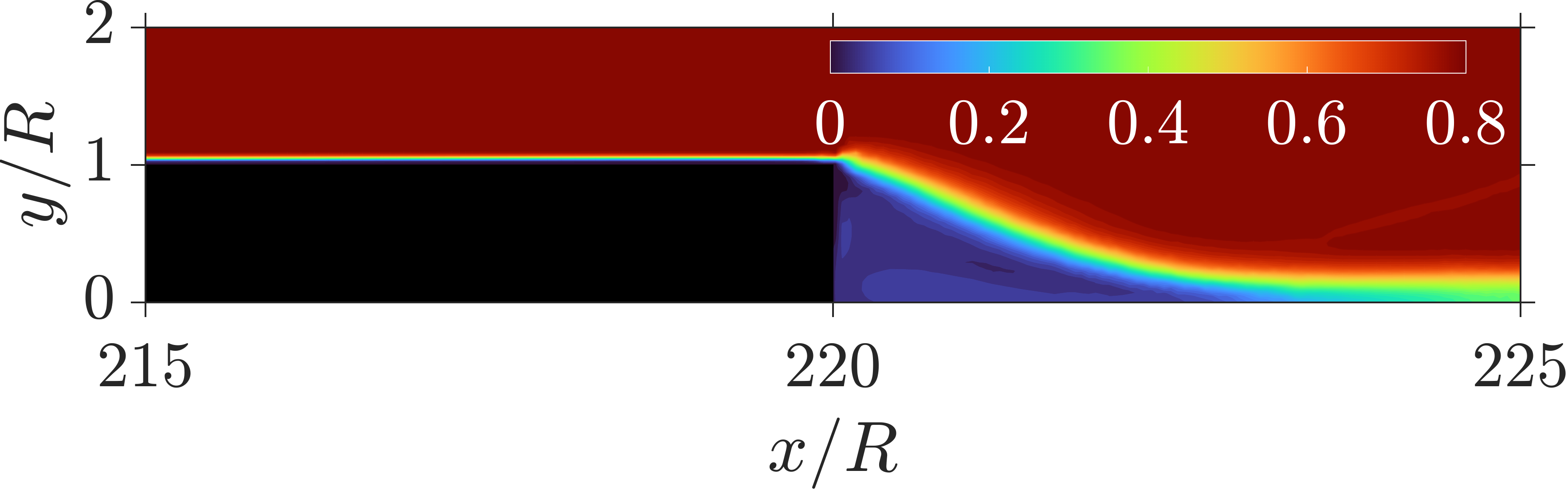}
    \caption{Contours of the metric of accuracy based on the the temperature approximation, $\mathcal{T}_i=\mathcal{T}_i(x,y,\alpha=0)$.}
    \label{fig:Ti contours}
\end{figure}

Based on the aforementioned error we introduce a metric of accuracy for the temperature approximation, $\mathcal{T}_i=\mathcal{T}_i(x,y,\alpha)$ for each coordinate in the 2D domain, $(x,y)$ and AoA, $\alpha$. This metric will be used to inform the network when the adiabatic temperature approximation is reliable, and when it is not. In Figure~\ref{fig:Ti contours} we display contours of $\mathcal{T}_i$ resulting from the computed error in Figure~\ref{fig:error contours}. Regions of excellent accuracy are given values of $\mathcal{T}_i=0.8$ (this is the highest value that $\mathcal{T}_i$ can take), while poor accuracy is denoted with values tending to zero. Such procedure is followed for all AoAs.

\subsubsection{Metric Generation for the temperature-Mach number relation}\label{subsubsec: Metric Generation for the temperature-Mach number relation}
\begin{figure}[htbp!]
    \centering
    \includegraphics[width=0.35\textwidth]{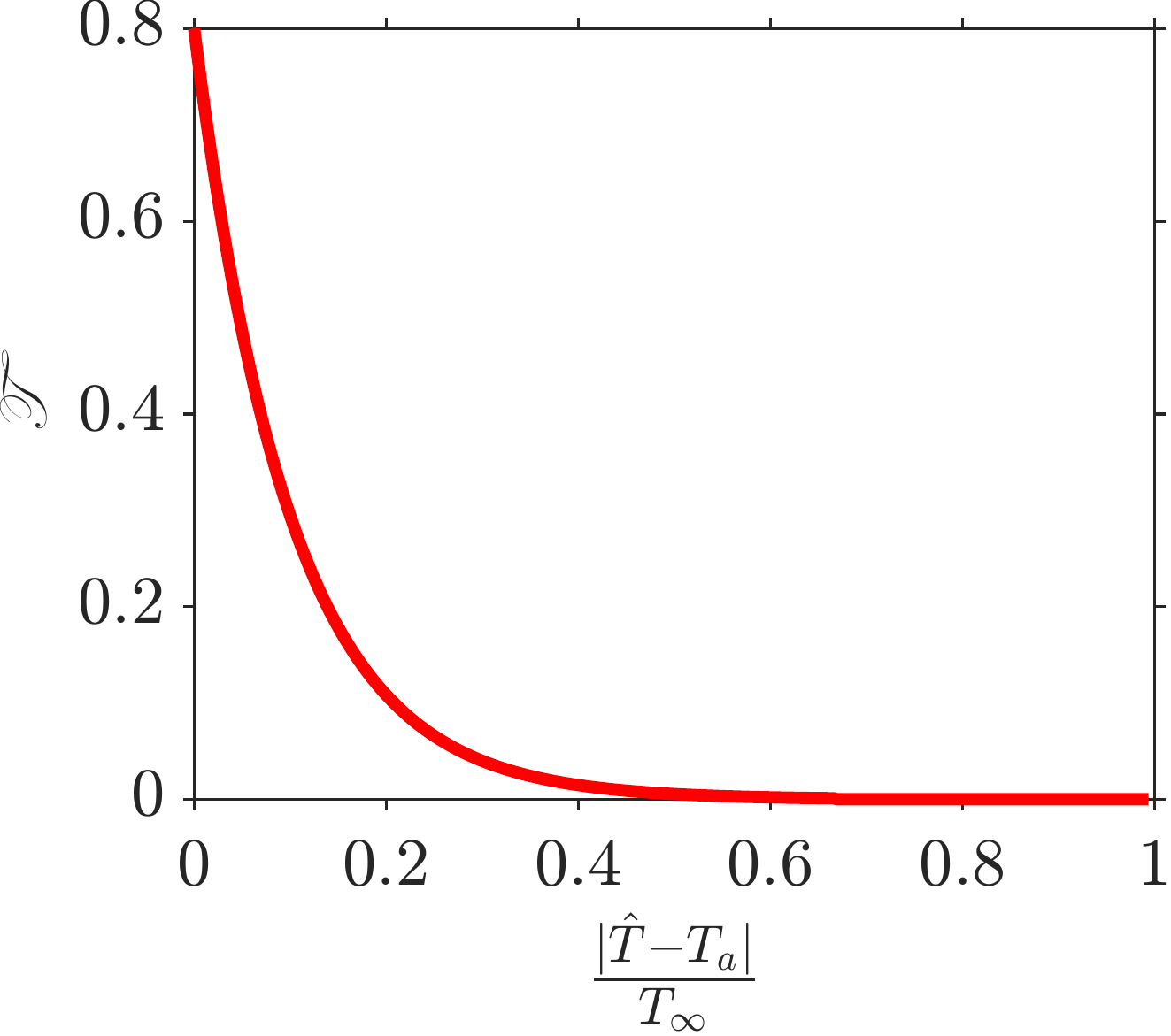}
    \caption{Metric of accuracy for the temperature approximation against the error $\varepsilon=\frac{|\hat{T} - T_a|}{T_{\infty}}$.}
    \label{fig:exponential}
\end{figure}

The relative difference between the CFD temperature data and the temperature approximation is accounted for in the error, 
\begin{equation}
    \varepsilon(x,y) = \frac{|\hat{T}(x,y) - T_a(x,y)|}{T_{\infty}},
\end{equation}
\noindent which is zero at highest accuracy of the approximation. The coefficient $\mathcal{T}_i$ is obtained directly from the computed error according to
\begin{equation}
    \mathcal{T}_i = 0.8 \exp(-k\varepsilon),
\end{equation}
\noindent such that when the error is nil $\mathcal{T}_{max}=0.8$ and at large error $\mathcal{T}_i$ drops exponentially to zero. The constant scalar is set to $k=10$ to heavily penalise the temperature approximation when the error is still moderate, as seen visually in figure \ref{fig:exponential}.

\subsubsection{Incorporating the temperature-Mach number relation into the networks}\label{subsubsec: Incorporating the temperature-Mach number relation into the networks}
To incorporate the temperature inductive bias into the network we modify the model temperature prediction layer using the following expression: 
\begin{equation}
    T=(1-\mathcal{T})T_d+\mathcal{T}T_a,
    \label{t_bc}
\end{equation}
where $T$ is the final temperature prediction, $T_d$ is the data-driven prediction, $T_a$ is the temperature approximation, and $\mathcal{T}$ the temperature approximation coefficient. Note that all of these are coordinate dependent. Also, since the maximum value for the temperature approximation coefficient is $\mathcal{T}_{max}=0.8$, there is always part of the solution influenced by the data-driven prediction $\tilde{T}$.

When both the adiabatic temperature approximation and the boundary condition enforcement for the temperature are combined, that is, putting together Equation~\ref{bc_temp} and Equation~\ref{t_bc}, we obtain:
\begin{equation}
    T = \left((1-\mathcal{T})T_d+\mathcal{T}T_a\right)(1-\lambda(\beta_3)w_i)+\lambda(\beta_3)T_0w_i,
    \label{total_eq}
\end{equation}
Note that everywhere off the wall Equation~\ref{total_eq} collapses to Equation~\ref{t_bc}.

\subsection{Models and Neural Network Architectures}\label{subsec: Models and Neural Network Architectures}
Next we give a summary of the model architectures used in this work. We use a baseline architecture and then extended incorporating the Fourier positional feature mappings, no-slip and isothermal wall boundary conditions enforcement, as well as the temperature-Mach number relation in regions of validity.

\subsubsection{Baseline Architecture}\label{subsubsec: Baseline Architecture}
The baseline model is a neural field, which in practice is an MLP, with 5 layers with hidden dimension of 50, one-dimensional batch normalization, and a Sigmoid-Weighted Linear Unit (SILU) activation function~\cite{Elfwing2017SigmoidWeightedLU}, in that order for each of the 5 layers. Note that the first layer has input dimension of 3, since it takes in the $x$ and $y$ coordinates, and $\alpha$. The latent representation learned by these layers is then passed on to 4 independent blocks of layers, each to predict one of the following aerothermodynamic quantities: pressure, temperature, and $x$- and $y$-velocity components. These blocks correspond to the $f_p,f_T,f_{v_x},f_{v_y}$ discussed in Section~\ref{subsec: Multi-Task Learning}. Each of the blocks is composed of a linear layer, a one-dimensional batch normalization, a SILU, another linear layer, a SILU, and a final linear layer. The output dimension for each of the blocks is 1 and the hidden dimension for all layers is 50, as before. As later discussed in Section~\ref{sec: Results}, the baseline model has very poor performance and using Fourier positional feature mappings is \textit{key} to obtain a good model.

\subsubsection{Modifications to the Baseline Model: Fourier Positional Feature Mappings and Physics Inductive Biases}\label{subsubsec: Modifications to the Baseline Model: Fourier Positional Feature Mappings and Physics Inductive Biases}
As previously mentioned in Section~\ref{subsec: Neural Fields and Fourier Positional Feature Mappings}, we incorporate Fourier positional feature mapping to improve the model performance. To do so, we use a projection matrix $\mathbf{B}\in\mathbb{R}^{m\times 2}$ and set $m=400$, which is chosen empirically. This effectively means that the input dimension of the first layer of the model must be modified to be $m\times2+1=801$, where the additional channel is still necessary for the AoA. We leave for future work exploring whether using feature mappings to encode the AoA would also benefit the model, but for now we simply input it as a scalar value. $\sigma^{2}$ is adjusted using a grid search and set to $\sigma^{2}=30^{2}$ for best performance. This is later discussed in Section~\ref{sec: Results}.

For the no-slip boundary condition enforcement and the temperature-Mach number relation no major changes to the baseline model are required. After the model blocks for the velocity components (equations~\ref{vx_bc} and~\ref{vy_bc}) are implemented, the prediction made by the temperature block is used in equation~\ref{t_bc}.

\subsubsection{Nomenclature}\label{subsubsec: Nomenclature}
In total we have 5 models, which we will compare in Section~\ref{sec: Results}. We will refer to the base model with no Fourier feature mappings as NF$^{*}$ and to the one with the mappings as NF, where NF stands for neural field. The NF$^{*}$ model is not able to capture the high-frequency fluctuations in the data, as later shown in Section~\ref{sec: Results}. Hence, we will implement all additional modifications to the NF model. We will refer to the modified model with no-slip and isothermal wall boundary condition enforcement as NF-BC (Equations~\ref{vx_bc},~\ref{vy_bc}, and~\ref{bc_temp}) and the model with the temperature-Mach number relation as NF-T (Equation~\ref{t_bc}). The model that combines both the aforementioned techniques will be named NF-BC-T. 

\subsubsection{Model Training}\label{subsubsec: Model Training}
We train the models using only two AoAs $$\alpha_{train} \in [0^{\circ},9^{\circ}],$$ and we test the model performance on $$\alpha_{test} \in [1^{\circ},2^{\circ},3^{\circ},4^{\circ},5^{\circ},6^{\circ},7^{\circ},8^{\circ},9.5^{\circ},10^{\circ},11^{\circ},11.5^{\circ},12^{\circ}],$$ which are reasonable operational AoAs. During testing, the models must be able to interpolate for lower AoAs as well as to extrapolate to higher AoAs. Also, note that we \textit{only} require two CFD snapshots of resolved flow around the rocket to train our model and achieve competitive performance. This showcases the data efficiency of our model and its potential to become a useful tool for future rocket design. We train all models for 10,000 epochs and with a learning rate $\eta = 10^{-3}$. We also normalize the data before training for better convergence. In particular, we normalize each of the aerothermodynamic quantities by its maximum value over the entire domain at each AoA, so that all values are at maximum 1.

\section{Results}\label{sec: Results}
In this section we discuss the results. First, we perform an in-depth aerothermodynamic analysis of the results, both qualitative and quantitative. After the aerothermodynamic study, we corroborate our findings using quantitative metrics from the machine learning literature. 

\subsection{Aerothermodynamic Study}\label{subsec: Aerothermodynamic Study}

First, in Section~\ref{A Preliminary Analysis of the Baseline Model and the Importance of Fourier Positional Feature Mappings} we start by analysying the performance of the NF$^{*}$ and NF models, which are purely data-driven models with no physics knowledge incorporated into them. Next, in Section~\ref{The Benefits of Incorporating Physics-based Inductive Biases} we analyze the benefits of encoding known aerothermodynamic physics into the baseline neural fields.

\subsubsection{A Preliminary Analysis of the Baseline Model and the Importance of Fourier Positional Feature Mappings}
\label{A Preliminary Analysis of the Baseline Model and the Importance of Fourier Positional Feature Mappings}
We investigate the predictive accuracy of the models introduced in Section~\ref{subsubsec: Nomenclature} by examining the critical flow features of the CFD test case. These include the supersonic bow shock located upstream of the body, the development of the velocity and thermal boundary layers, and the geometry of the recirculation zone located behind the body. Collectively, these flow characteristics contribute to predicting the aerothermal performance of the geometry being studied.

\begin{figure}[htbp!]
\centering
\begin{subfigure}
  \centering
  \includegraphics[width=\linewidth]{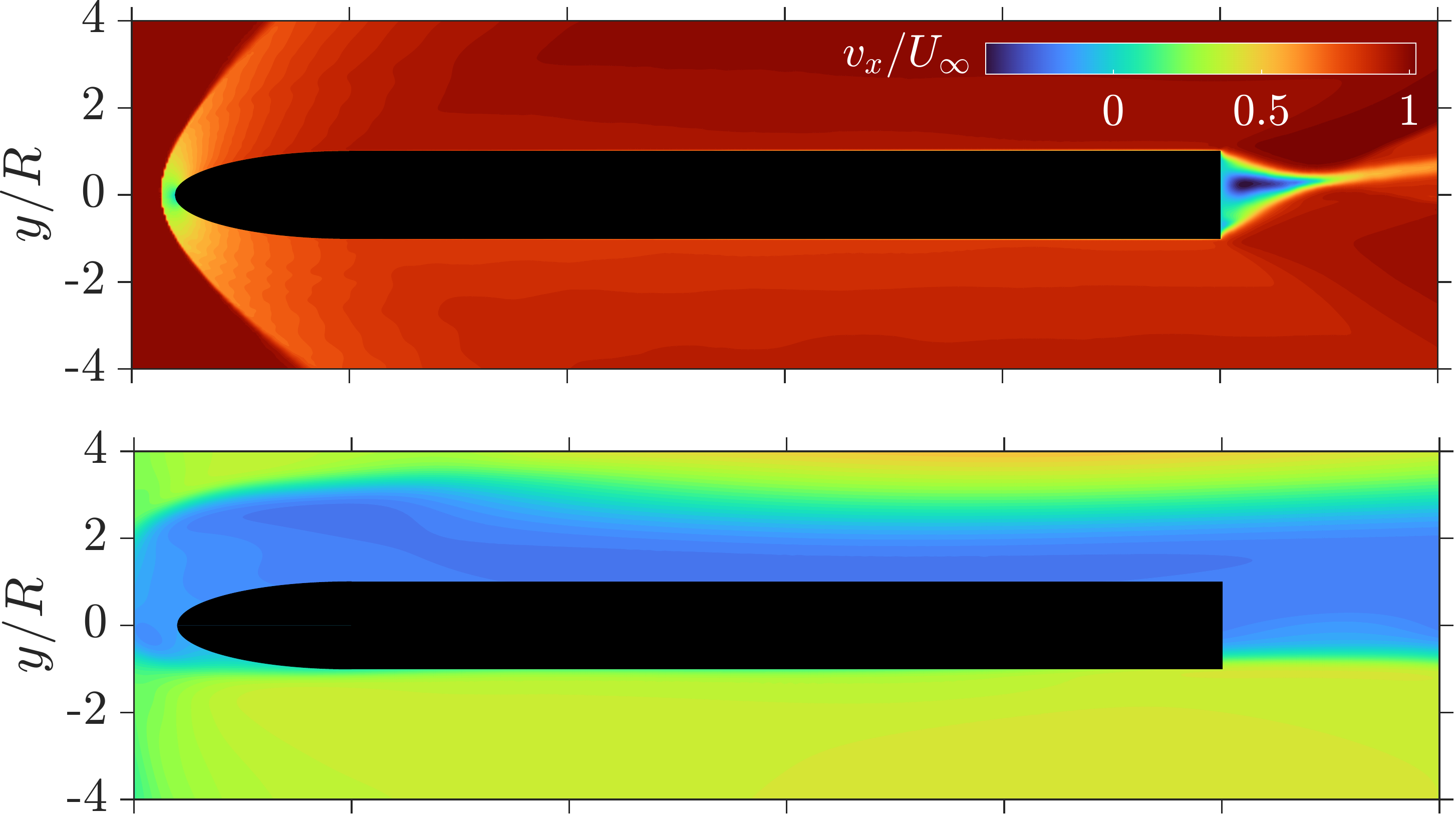}
\end{subfigure}
\begin{subfigure}
  \centering
  \includegraphics[width=1.015\linewidth]{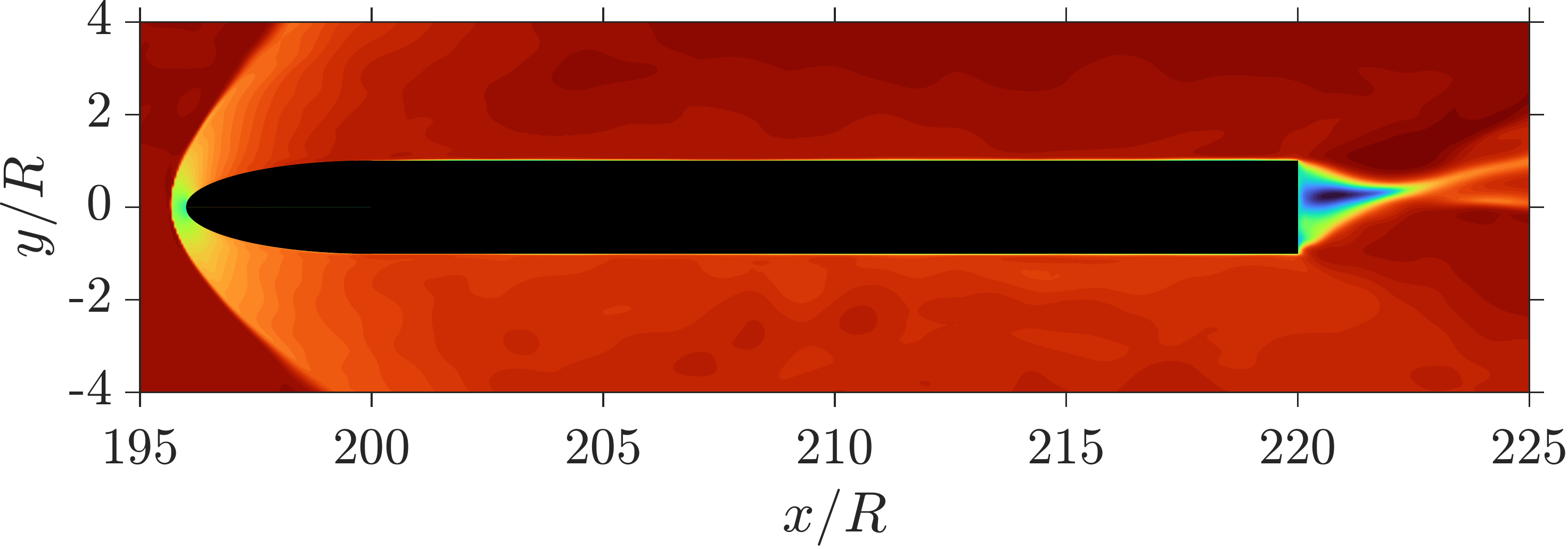}
\end{subfigure}
\caption{Comparison of contours of normalised $v_x$ velocity from (top) ground truth, (middle) NF$^{*}$, and (bottom) NF models at $\alpha=6^{\circ}$.}
\label{fig:noencoding_NF_f30_vx_contours_AoA6}
\end{figure}

Figure~\ref{fig:noencoding_NF_f30_vx_contours_AoA6} compares the normalised horizontal velocity field, $v_x/U_{\infty}$, of the NF$^{*}$ model against the NF model at $6^{\circ}$ incidence. Given that the models are trained on $\alpha_{train}~\in~ [0^{\circ},9^{\circ}]$, this test case incidence requires the model being able to interpolate between previously seen AoAs. 
Unarguably, the NF$^{*}$ model (middle panel) fails to capture the flow topology in every aspect, showing large nonphysical negative velocity areas above the body and missing the bow shock and wake features completely. On the other hand, the NF model (bottom panel) shows remarkable improvements, suggesting the fundamental role of Fourier positional feature mappings. Both the shock front and the wake are resolved with minor discrepancies. In particular, the wake is bent slightly upwards compared to the ground truth. We also observe small background fluctuations appearing as ragged contours in the prediction. 

\begin{figure}[htbp!]
    \centering
    \includegraphics[width=0.475\textwidth]{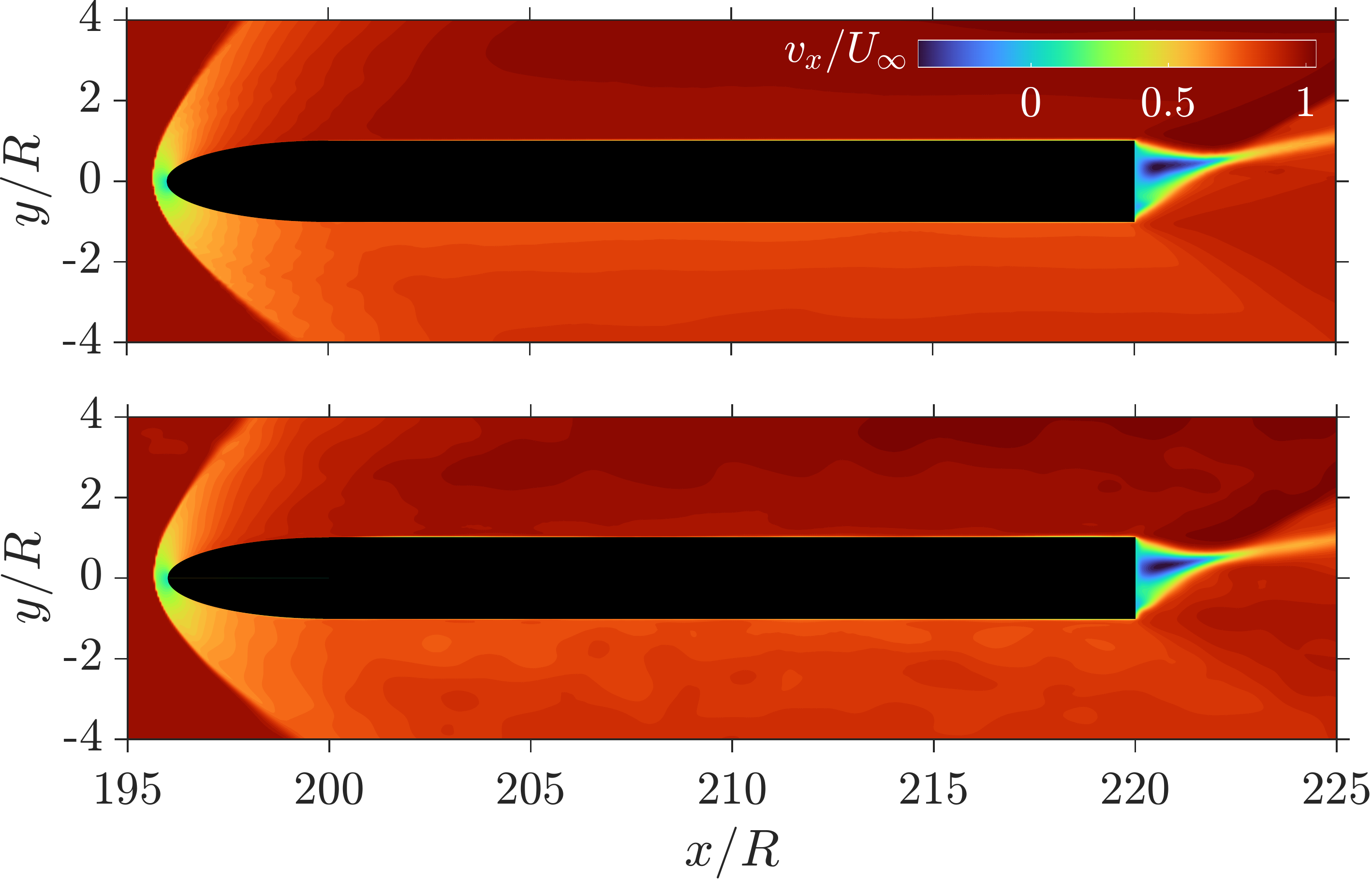}
    \caption{Contours of normalised $v_x$ velocity from the NF model at $\alpha=10^{\circ}$. Truth: top. Prediction: bottom.}
    \label{fig:NF_f30_vx_contours_AoA10}
\end{figure}

To assess the quality of flow reconstruction of the NF model in the extrapolation AoA range, we plot the same contours in figure~\ref{fig:NF_f30_vx_contours_AoA10} at $\alpha=10^{\circ}$. Overall, we notice similar accuracy in the shock region but improved match with respect to the true data in the wake. Although $\alpha=10^{\circ}$ involves extrapolation, this AoA is closer to the training data than $\alpha=6^{\circ}$ in figure~\ref{fig:noencoding_NF_f30_vx_contours_AoA6}, which explains the improved performance. In this case we also notice a small discrepancy in the main direction of the wake, however the predicted wake has a smaller upward inclination compared to the ground truth. We identify the vortex core of the recirculating zone (dark blue contours) for both prediction and truth and notice that in the former the peak reverse-flow is $\approx40\% U_{\infty}$ at $y/R \approx 0.30$, and in the latter $\approx41\% U_{\infty}$ at $y/R \approx 0.35$. These locations confirm the small discrepancy in the wake inclination.

\begin{figure}[htbp!]
    \centering
    \includegraphics[width=0.4\textwidth]{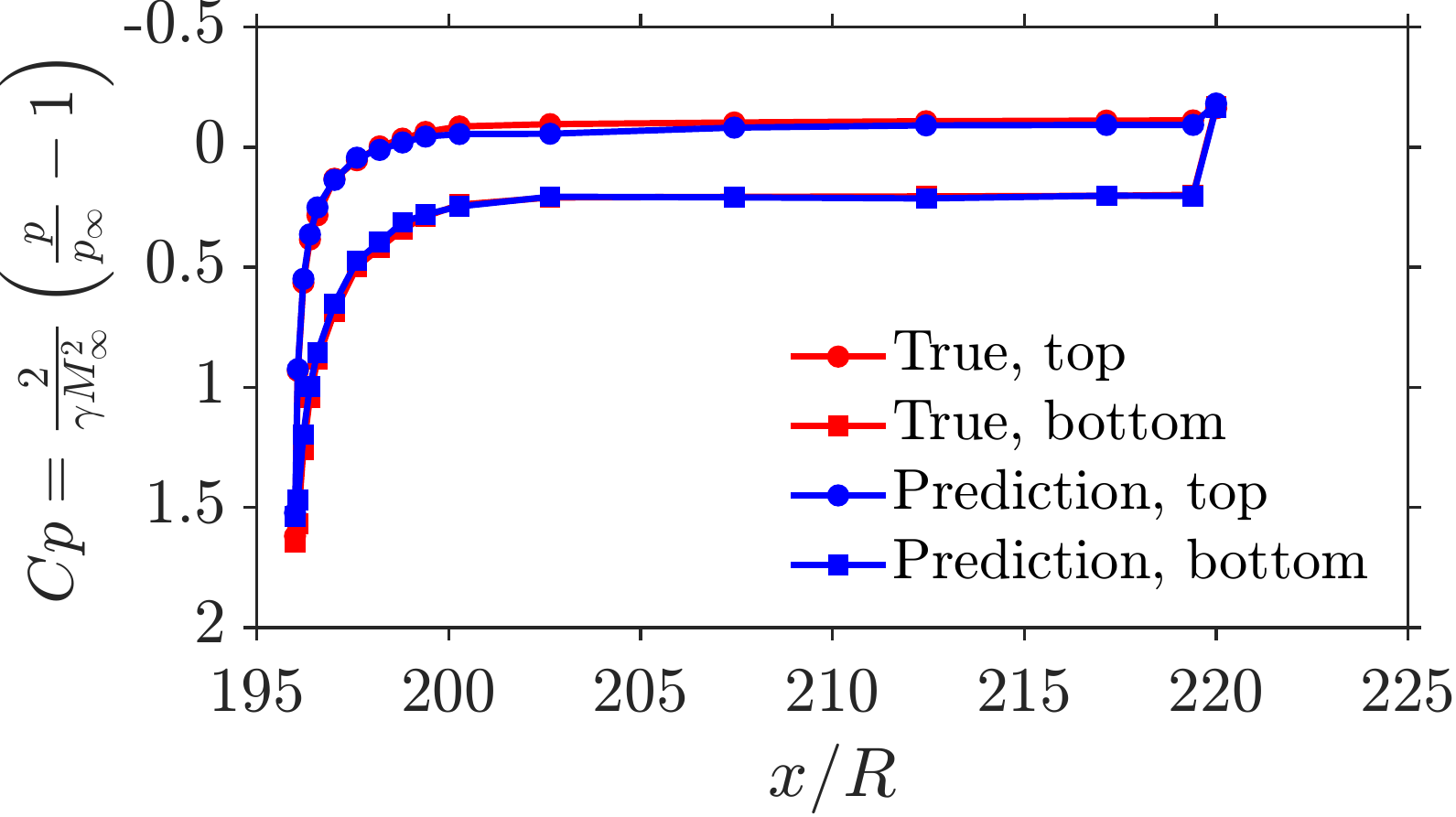}
    \caption{Pressure coefficient along the streamwise direction from the NF model at $\alpha=10^{\circ}$.}
    \label{fig:NF_f30_Cp_AoA10}
\end{figure}

For the same model architecture and AoA case we compute the pressure coefficient distribution around the body in figure~\ref{fig:NF_f30_Cp_AoA10}. 
Overall, we see great agreement with the true data. The location of the $C_{p,max}$ point is predicted spot-on at $x/R\approx196$, with the truth being 1.65 and the prediction 1.54 ($\approx6.7\%$ difference). 
Furthermore, the acceleration of the flow around the curved nose geometry is captured extremely well by the model, as well as the pressure plateau region on both top and bottom surfaces. Another discrepancy in the prediction is found in the value of back pressure, which is $-0.16$ compared to $-0.17$ in the truth (less than $6\%$ difference). This result confirms the ability of the model to learn the pressure field around the body, which is an important result for aerodynamic lift and drag performance analysis.

\subsubsection{The Benefits of Incorporating Physics-based Inductive Biases}
\label{The Benefits of Incorporating Physics-based Inductive Biases}

\begin{figure}[htbp!]
\centering
\begin{subfigure}
  \centering
  \includegraphics[width=\linewidth]{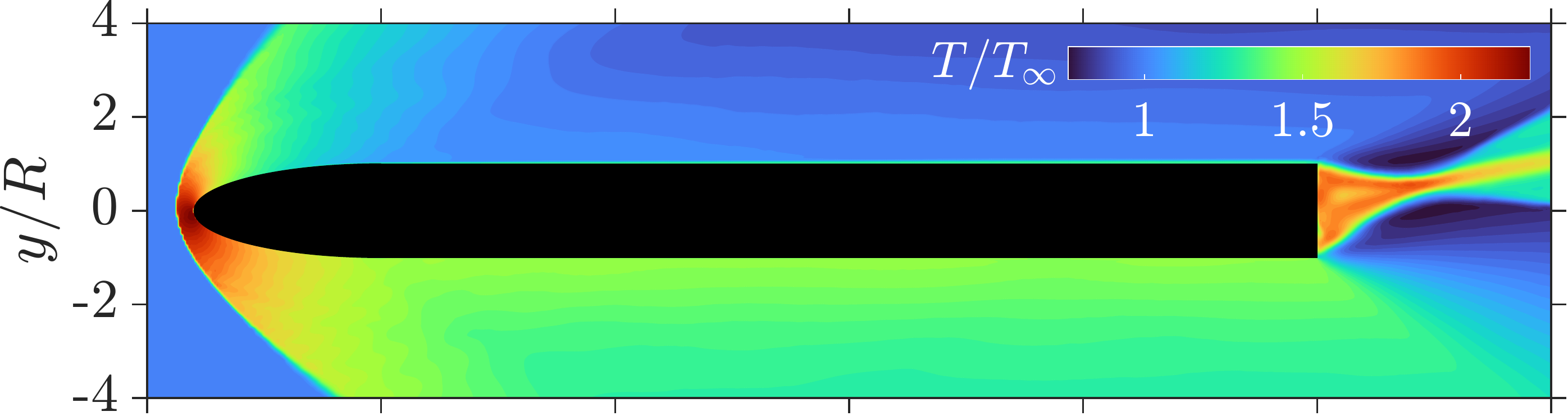}
\end{subfigure}
\begin{subfigure}
  \centering
  \includegraphics[width=\linewidth]{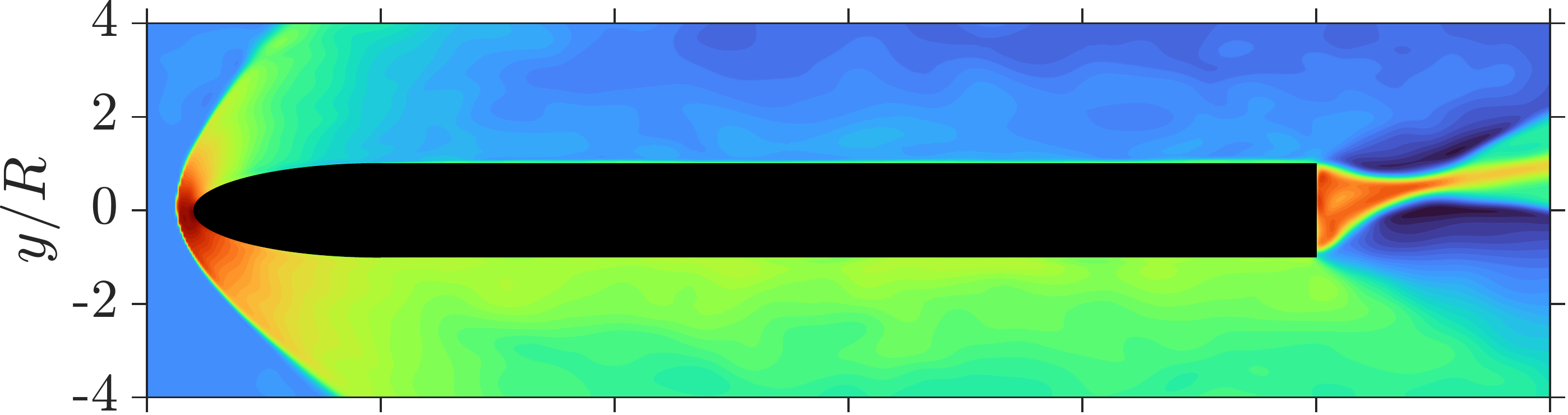}
\end{subfigure}
\begin{subfigure}
  \centering
  \includegraphics[width=1.015\linewidth]{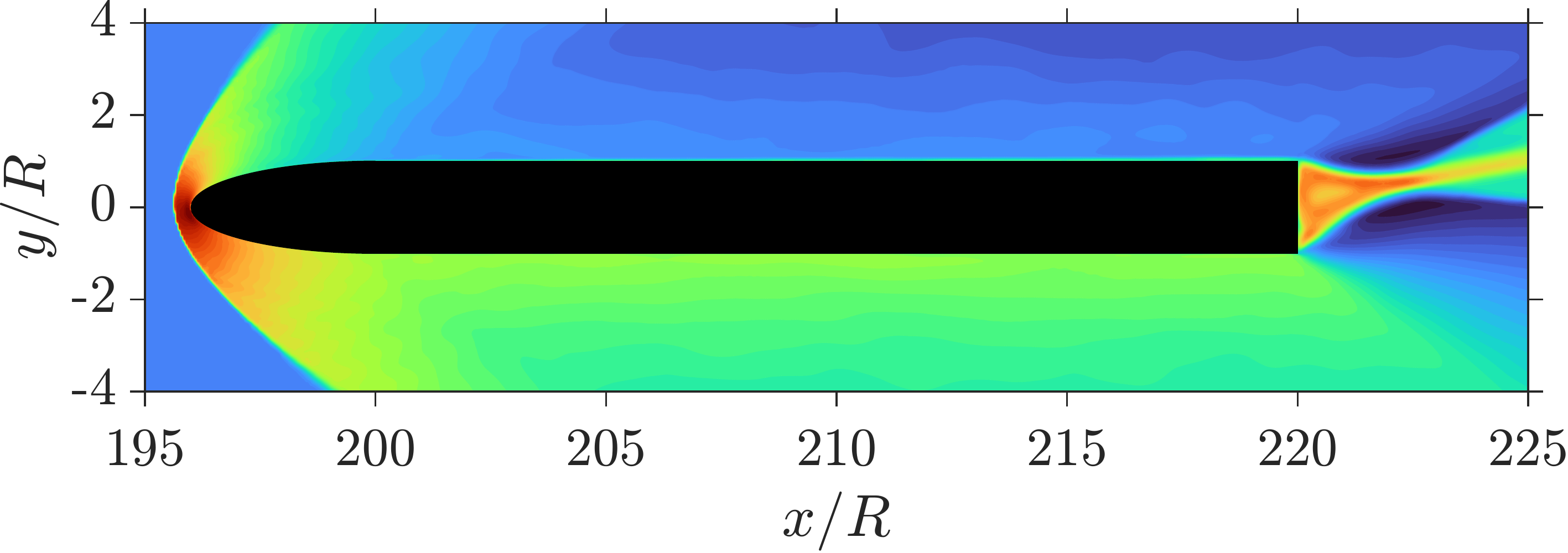}
\end{subfigure}
\caption{Comparison of contours of normalised temperature from (top) ground truth, (middle) NF and (bottom) NF-T models at $\alpha=10^{\circ}$.}
\label{fig:NFandNFT_T_contours_AoA10}
\end{figure}

To achieve faster learning and incorporate physical information into the model, we add the temperature-Mach relation bias as explained in \ref{subsubsec: Incorporating the temperature-Mach number relation into the networks}. Figure~\ref{fig:NFandNFT_T_contours_AoA10} displays temperature contours that compare the accuracy of the NF and NF-T models. Overall, the adiabatic temperature relation reduces spurious fluctuations in the predicted solution and improves the matching of the wake inclination angle. The NF-T model accurately predicts the $T_{max}$ value, which matches the true data exactly, with a value of $2.24T_{\infty}$. Conversely, the NF model gives a value of $2.19T_{\infty}$. Generally, the adiabatic temperature relation provides a better representation of the high-temperature post-shock flow region.

\begin{figure}[htbp!]
    \centering
    \includegraphics[width=0.45\textwidth]{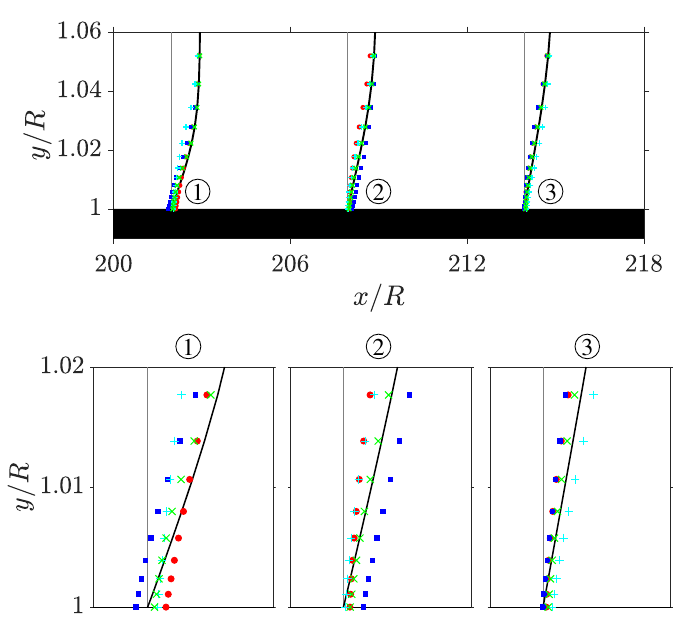}
    \caption{Comparison of horizontal velocity $v_x$ boundary layer profiles from (\textcolor{red}{$\bullet$}) NF, (\textcolor{blue}{$\blacksquare$}) NF-BC, (\textcolor{cyan}{$\mathbf{+}$}) NF-T and (\textcolor{green}{$\mathbf{\times}$}) NF-BC-T models at 3 prescribed $x$-locations on the upper surface at $\alpha=10^{\circ}$. The truth is plotted with black-solid lines, while the zero velocity at the wall is indicated by grey-solid lines. The coordinate $y/R=1$ marks the location of the wall.}
    \label{fig:All_vxBL_AoA10}
\end{figure}

\begin{figure}[htbp!]
    \centering
    \includegraphics[width=0.45\textwidth]{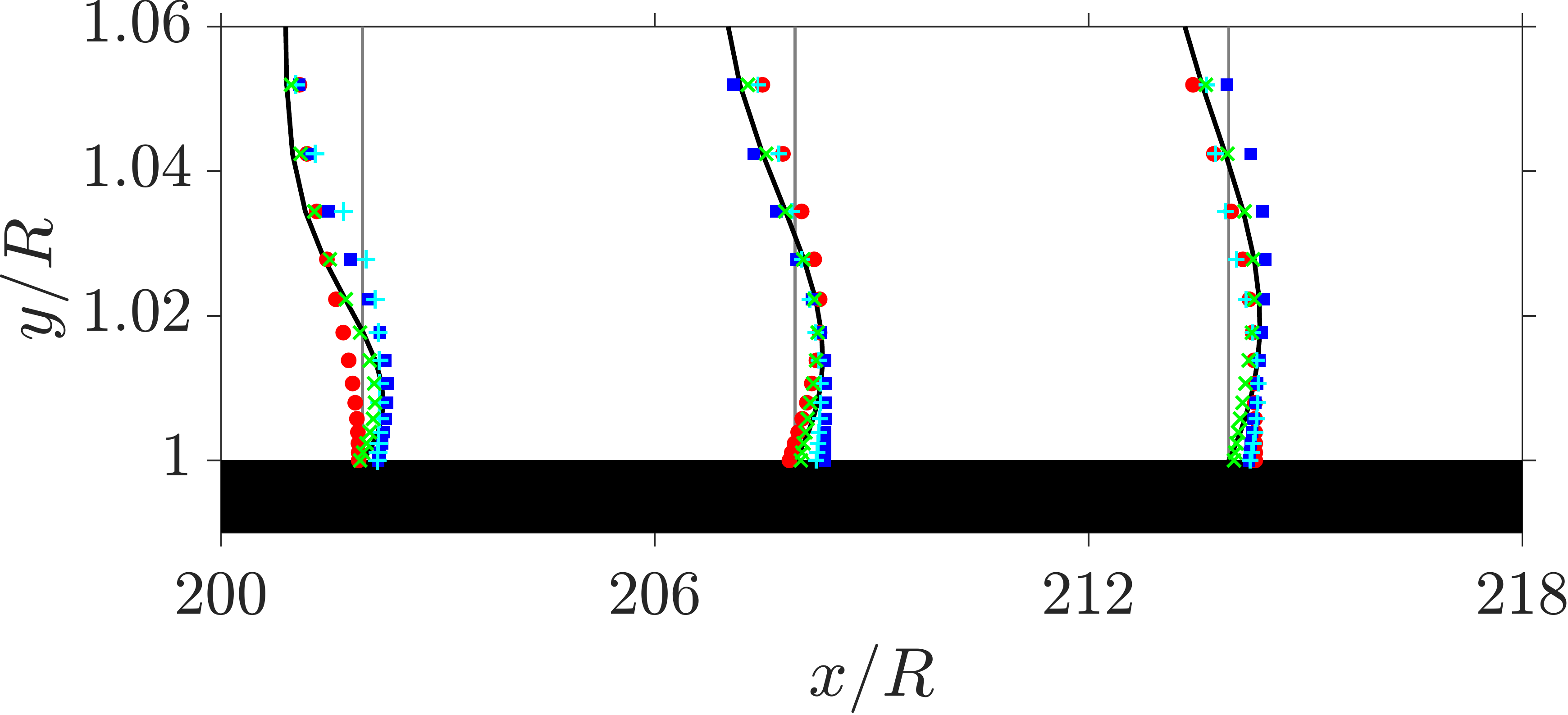}
    \caption{Comparison of temperature $T$ boundary layer profiles from (\textcolor{red}{$\bullet$}) NF, (\textcolor{blue}{$\blacksquare$}) NF-BC, (\textcolor{cyan}{$\mathbf{+}$}) NF-T and (\textcolor{green}{$\mathbf{\times}$}) NF-BC-T models at 3 prescribed $x$-locations on the upper surface at $\alpha=10^{\circ}$. The truth is plotted with black-solid lines, while the wall temperature, $T_{wall}$, is indicated by grey-solid lines. The coordinate $y/R=1$ marks the location of the wall.}
    \label{fig:All_TBL_AoA10}
\end{figure}

A closer look at the boundary layer profile is attempted at 3 discrete $x$-locations on the upper surface of the body, $x/R=[202,208,214]$. In figures~\ref{fig:All_vxBL_AoA10} and \ref{fig:All_TBL_AoA10} we visualise the $v_x$ and $T$ wall-normal profiles near the wall, located at coordinate $y/R=1$. In these two figures all of the four models (excluding NF$^{*}$) are compared. From the velocity profiles we see that all models eventually tend to the true data away from the wall, however remarkable differences are identified in the proximity of the wall. In general we see that the highest accuracy is achieved by the NF-BC-T model including all of the available physical information. 

While the accuracy changes across the three stations under consideration, we can safely state that the temperature bias has a significant impact on accuracy, not only for the reasons explained earlier in figure~\ref{fig:NFandNFT_T_contours_AoA10}, but also for the satisfaction of the wall temperature boundary condition. However, this condition is closely met when the wall temperature, $T_{wall}$, is enforced in the model. From both velocity and temperature profiles we also learn that the NF-BC model does not improve the prediction of the baseline NF model by any means, but does so only when we add the adiabatic temperature bias. The underlying reasons are still unclear to the authors, but, overall, we appreciate the importance of adding known physics to the architecture to restrict the solution space and obtain better predictive capability. The synergistic presence of these additional constraints seem to be beneficial to the model.


\subsubsection{Additional Analysis for the Best Performing Neural Field Model}

\begin{figure}[htbp!]
    \centering
    \includegraphics[width=0.45\textwidth]{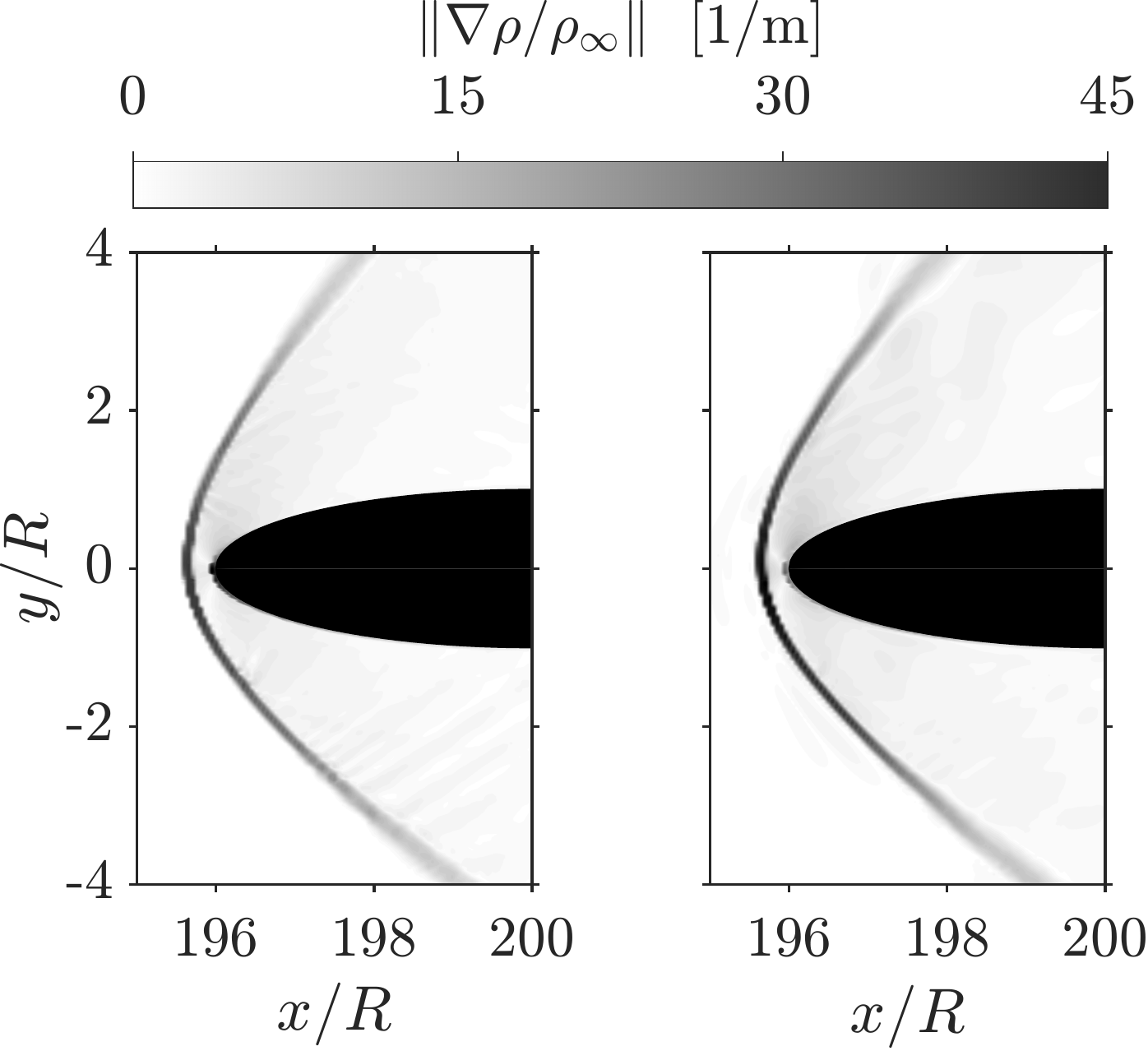}
    \caption{Contours of the magnitude of the normalised density gradient $\nabla \rho/\rho_{\infty}$ from the NF-BC-T model at $\alpha=10^{\circ}$. Truth: left. Prediction: right.}
    \label{fig:NFBCT_schlieren_AoA10}
\end{figure}

Lastly, we provide some further analysis for comparing the NF-BC-T model to the ground truth data. In figure~\ref{fig:NFBCT_schlieren_AoA10} we zoom in the bow shock region around the nose of the body and compute the magnitude of the gradient of the normalised density given by $\left \| \nabla \rho/\rho_{\infty} \right \| = \sqrt{\left ( \frac{\partial \rho/\rho_{\infty}}{\partial x}\right )^2 + \left ( \frac{\partial \rho/\rho_{\infty}}{\partial y} \right )^2}$. Immediately, we see that the shock profiles match well qualitatively. From a more quantitative viewpoint we infer the (non-dimensional) shock stand-off distance and asymptotic shock angle from the figure and obtain $0.345$ and $36.4^{\circ}$ from the ground truth data and $0.327$ and $37.7^{\circ}$ from the model at $\alpha=10^{\circ}$.

\begin{figure}
\centering
\begin{subfigure}
  \centering
  \includegraphics[width=0.8\linewidth]{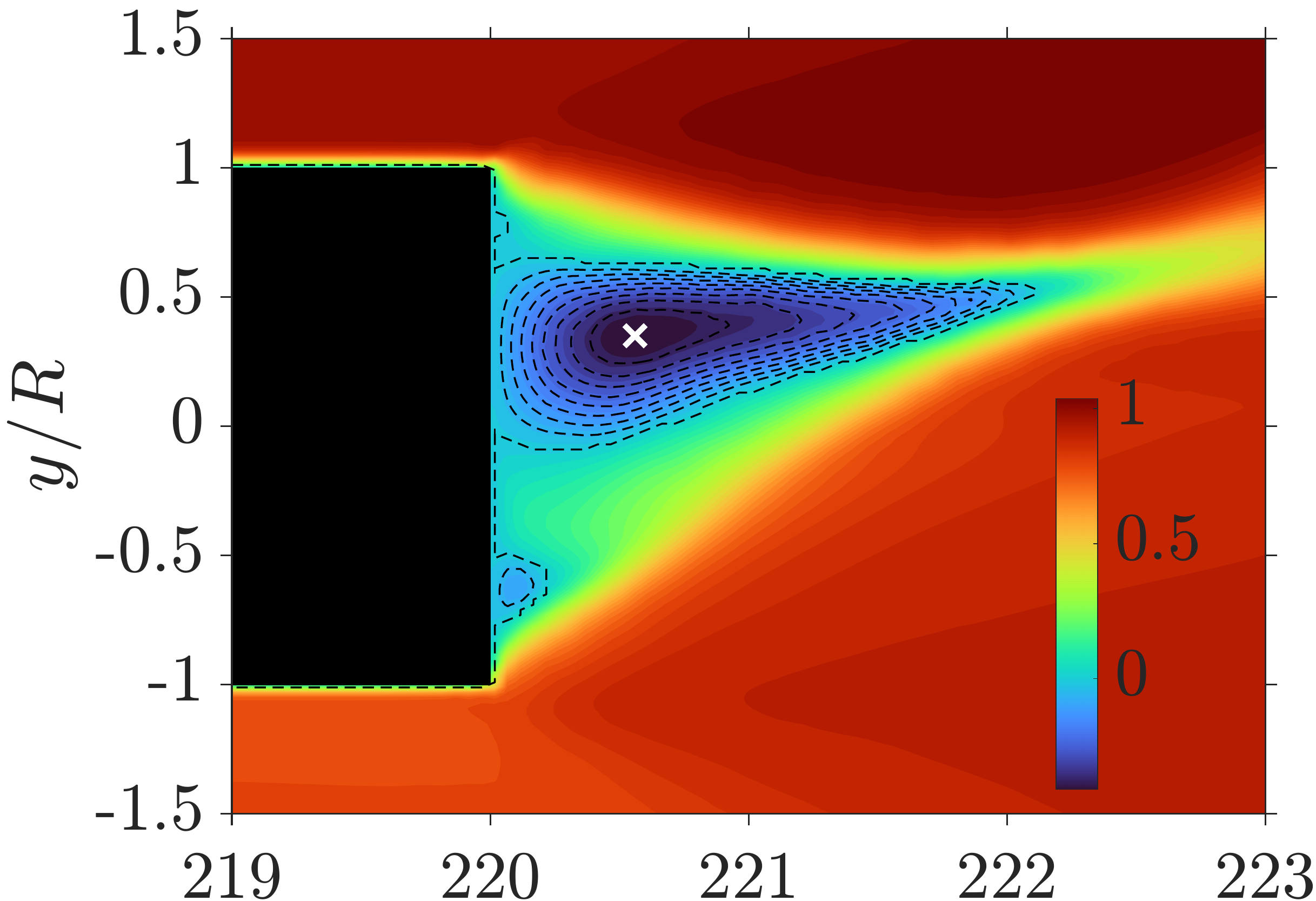}
\end{subfigure}
\begin{subfigure}
  \centering
  \includegraphics[width=0.8\linewidth]{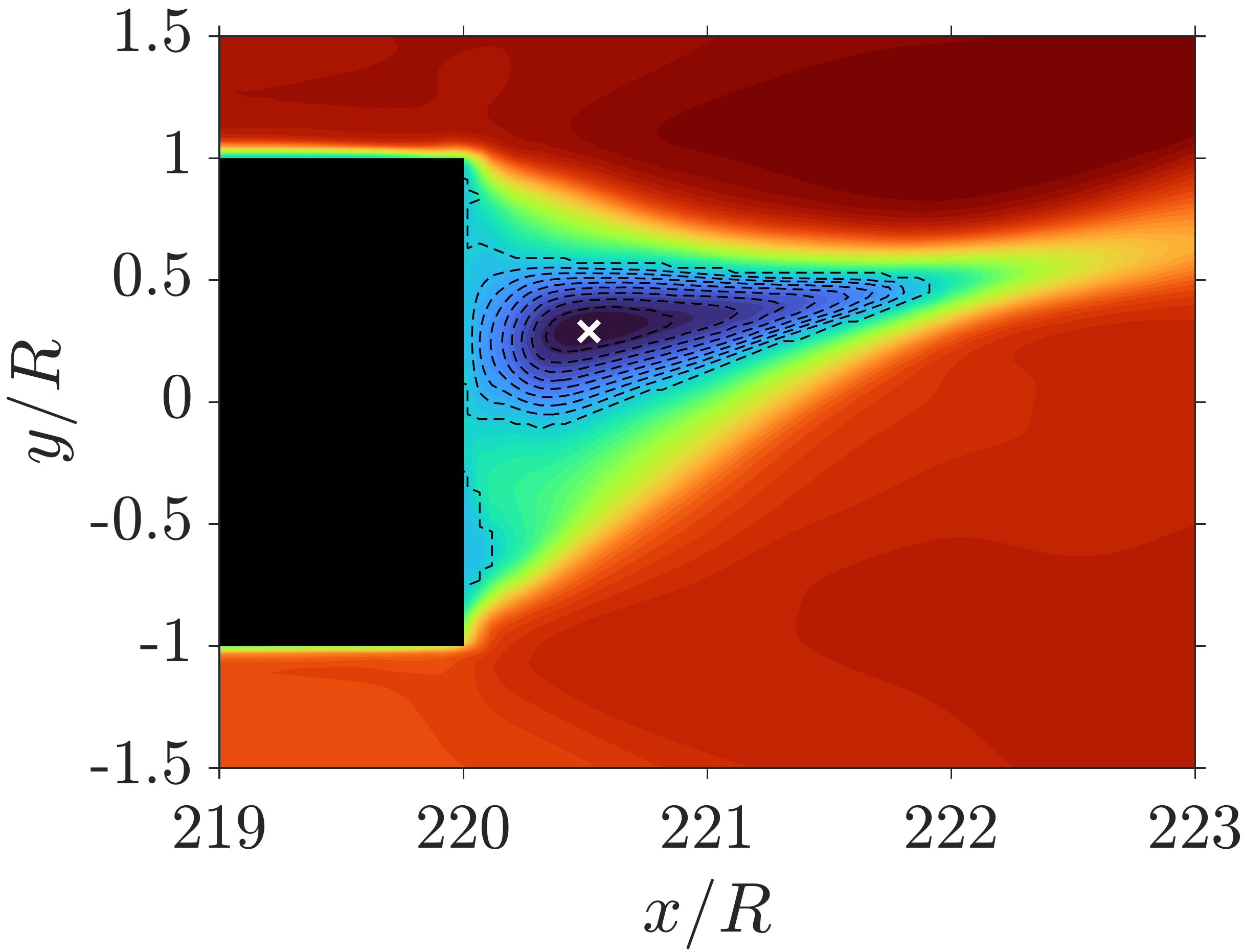}
\end{subfigure}
\caption{Normalised streamwise velocity field, $v_x/U_{\infty}$, in the wake at $\alpha=10^{\circ}$. Truth: top. Prediction: bottom (NF-BC-T model). Isolines of constant velocity are plotted using black-dashed lines. White cross markers denote the location of the vortex core.}
\label{fig:NFBCT_wake_AoA10}
\end{figure}

In figure~\ref{fig:NFBCT_wake_AoA10} we compare the isolines of constant streamwise velocity. The vortex structure agrees both qualitatively and quantitatively, with the vortex core located at $[x,y]=[220.56R,0.35R]$ and $[x,y]=[220.526R,0.30R]$ for the true data and the NF-BC-T model, respectively.

\begin{figure}
    \centering
    \includegraphics[width=0.4\textwidth]{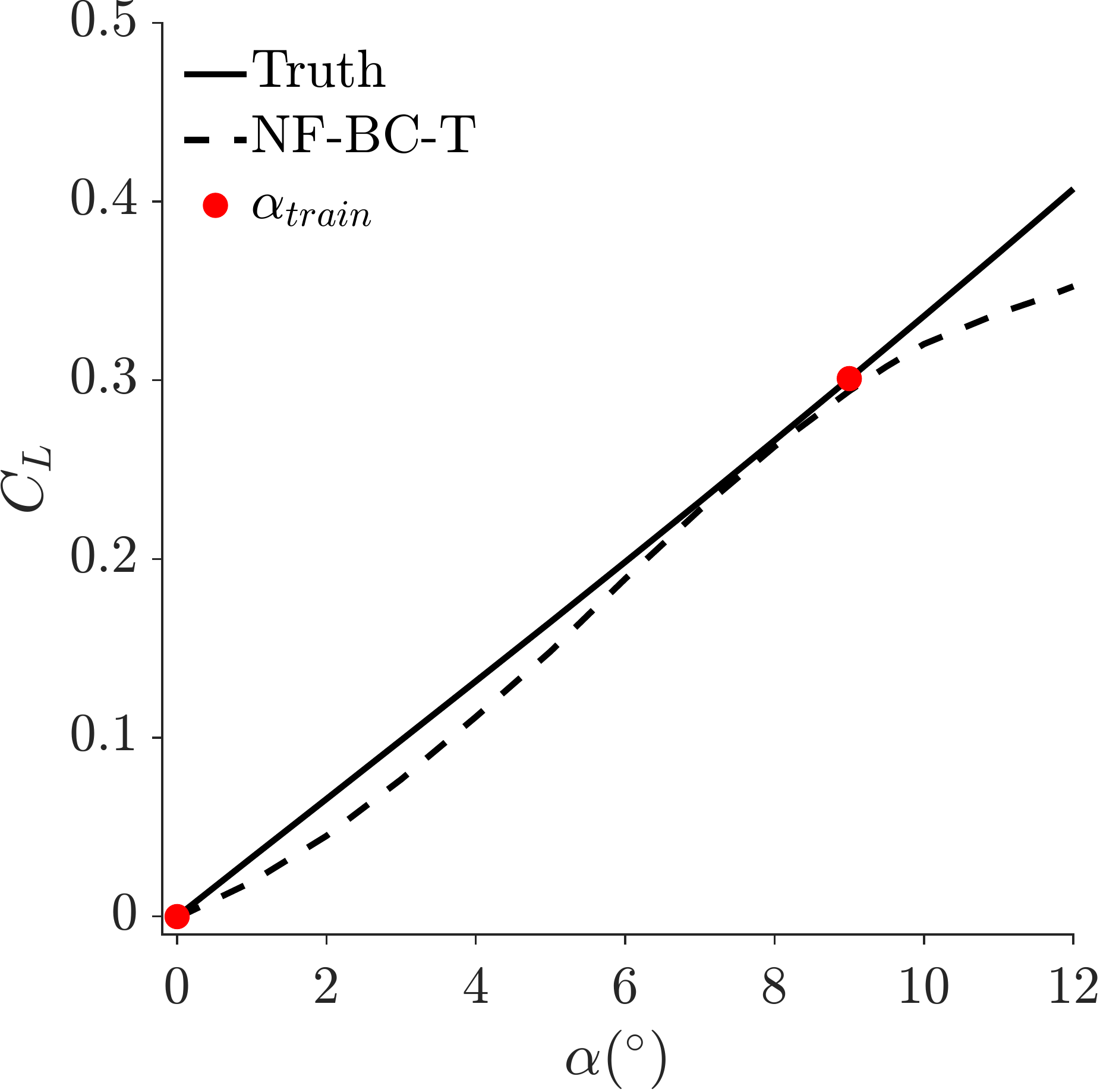}
    \caption{Lift coefficient curve predicted by the NF-BC-T model compared with CFD data. The red circles denote the two angles of attack used for training, $\alpha_{train} \in [0^{\circ},9^{\circ}]$.}
    \label{fig:NFBCT_CL}
\end{figure}

As a final comparison of the best identified model (NF-BC-T) with the true data, figure~\ref{fig:NFBCT_CL} shows the lift coefficient given by the CFD simulations (truth) and the trained model. The AoAs used for training are also highlighted in the plot. As expected, the model predicts the true data best close to the training AoAs, i.e. $\alpha_{train} \in [0^{\circ},9^{\circ}]$. Reasonable predictions are obtained in the interpolated interval $0^{\circ} \leq \alpha \leq 9^{\circ}$, with a maximum error of 30\% at $\alpha = 2.5^{\circ}$. It is noted that for angles higher then $9^{\circ}$, the lift coefficient predicted by the model starts diverging from the true data, indicating a worse performance for the model when trying to extrapolate at higher AoAs rather than when interpolating between the trained AoAs. 

\subsection{Machine Learning Performance Metrics}\label{subsec: Machine Learning Performance Metrics}
In order to assess the predictive capabilities of the ROM, we employ several performance indicators from the machine learning literature. The objective is to support the aerothermodynamics analysis with measurable outcomes. To obtain a more comprehensive perspective, we will present the findings using multiple indicators as distinct metrics may overlook certain patterns or behaviors in the data.

\textbf{PSNR.} The Peak Signal-to-Noise Ratio~(PSNR) is a widely used metric for evaluating the quality of compressed or reconstructed digital images and videos in the neural field literature~\cite{Mildenhall2020NeRFRS,Xie2021NeuralFI}. In our case, we can consider the reconstructed from to be our "image". PSNR measures the difference between an original image or video and its reconstructed version, by quantifying the level of noise or distortion introduced during the reconstruction process. A higher PSNR value indicates better quality, with less distortion or noise. Typically, a PSNR value of 30 dB or higher is considered to be acceptable for most applications, while a PSNR value of 40 dB or higher is considered to be very good quality. 

\textbf{SSIM.} The Structural Similarity Index Measure~(SSIM) is another widely used metric for evaluating the quality of compressed or reconstructed digital images and videos~\cite{Wang2004ImageQA}, particularly in situations where human perception of quality is important. Unlike PSNR, which is based on pixel-wise differences between the original and reconstructed images, SSIM takes into account the structural similarities and differences between the images. Specifically, SSIM measures the similarity between two images by comparing their luminance, contrast, and structural information, which are known to be important factors in human visual perception. The SSIM value is a decimal between 0 and 1, with 1 indicating perfect similarity between the images.

\textbf{R$^{2}$.} The R$^{2}$ score, also known as the coefficient of determination, is a commonly used metric in machine learning and statistics to evaluate the performance of a regression model. It measures the proportion of the variance in the target variable that can be explained by the regression model. The R$^{2}$ score takes values between 0 and 1, with 1 indicating a perfect fit between the model's predictions and the actual values of the target variable, and 0 indicating that the model does not explain any of the variance in the target variable.

In the case of PSNR and SSIM we apply off-the-shelf functions to evaluate the model performance. However, for R$^{2}$ we apply a slight variation considering the formulation, 
\begin{equation}
    \mathrm{R}^{2} = \frac{\sum \left ( \hat{\phi}\left ( x,y \right ) - \phi\left ( x,y \right )\right )^2 }{\sum \left ( \hat{\phi}\left ( x,y \right ) - \phi_{\infty}\right )^2 },
    \label{eq:R2}
\end{equation}
where $\hat{\phi}$, $\phi$ and $\phi_{\infty}=[p_{\infty},T_{\infty},U_{\infty}\cos\alpha,U_{\infty}\sin\alpha]$ denote the true, the predicted variable and the value at free stream conditions, respectively. We also compute the R$^{2}$ score in a restricted region of the domain of size $x/R \in [195,225]$, $y/R \in [-1,1]$ to evaluate the accuracy of the models in the vicinity of the body, where high gradients in the field variables are located. This will be denoted as R$^{2}_{r}$.

\subsection{Quantitative Evaluation}\label{subsec: Quantitative Evaluation}
Lastly, we corroborate the finding presented in Section~\ref{subsec: Aerothermodynamic Study}. Table~\ref{tab:quant_results} displays the results using different metrics for all models. In line with the previous aerothermodynamics analysis, we can see that the NF$^{*}$ model gives very poor performance in terms of all metrics. Incorporating Fourier positional feature mappings boosts the scores substantially, for example, the test set pressure prediction using the NF model is almost three times higher than that using NF$^{*}$ considering PSNR, and the improvement is even more dramatic if we look at the R$^{2}$ score. The R$^{2}_{r}$ metric shows the same trend as the R$^{2}$, with consistently slightly higher values. This is expected since the original CFD data has higher resolution near the rocket. 

Adding physics to the model, such as the no-slip boundary condition and the adiabatic temperature approximation helps improve performance even further. From a quantitative perspective, we can see that the NF-BC-T consistently outperforms all other models, which agrees with the findings from Section~\ref{subsec: Aerothermodynamic Study}.

When comparing the aerothermodynamics quantities of interest in this model - pressure, temperature, and velocities - we observe that the NF-BC-T model achieves the best results for temperature, which is expected as the model was provided with substantial temperature inductive biases. However, the vertical velocity component performs relatively poorly for all models, which aligns with the conventional understanding that predicting this quantity accurately is challenging. The vertical component is influenced by several complex and interconnected factors, such as gravity, lift and drag, and atmospheric conditions like wind shear, turbulence, and temperature gradients. Although the dataset in this case is not overly complicated, it is evident from the results in Table~\ref{tab:quant_results} that the neural fields face significant difficulty with this component.

\begin{table}[hbtp!]
    \centering
    \caption{Quantitative results, mean and standard deviation using different evaluation metrics. Best scores are highlighted in \textcolor{red}{red}.}
    \include{tables/table}
    \label{tab:quant_results}
\end{table}

\section{Conclusion}\label{sec: Conclusion}
A predictive ROM based on a neural field was developed by simulating the laminar, steady flow around a simplified 2D rocket geometry at Mach 2.5 using CFD at different angles of attack. The proposed model architecture is capable of reconstructing the main flow features and interpolating continuously along the solution space spanned by the varying AoA parameter used in this study. As a result, a continuous aerothermal proxy model was developed to study the aerothermodynamics of a simplified rocket body.

The baseline model was enhanced with Fourier positional feature mappings and physical information. The former significantly improved the accuracy of the reconstructed flow fields compared to the basic neural field, NF$^{*}$, enabling prediction of the primary flow features such as the bow shock and wake profiles, as well as the pressure distribution around the body. The latter allowed for greater accuracy in the boundary layer region, especially with the NF-BC-T model which enforced no-slip and wall temperature boundary conditions at the body's surfaces and introduced an inductive bias for the temperature field based on the adiabatic temperature-Mach relation.

Neural fields offer several advantages over traditional CFD simulations in aerodynamics. They can interpolate quickly and efficiently in terms of AoA, providing accurate results in seconds compared to the time required for CFD simulations. This speed is especially useful in design iterations and optimization. Neural fields also generate continuous interpolations, unlike traditional methods that only provide discrete solutions. This is beneficial when analyzing complex aerodynamic systems where the flow field can exhibit gradual changes. Future extensions of this work could include a 3D test case and different aerodynamic shapes, as well as other choices of varying parameters such as the nose geometry, flight Mach number, and Reynolds number. Also, the model presented here could be trained on higher AoAs so that it could predict separation.

\bibliographystyle{plain}
\bibliography{references}
\end{document}

%% file: orcid.tex
\usepackage{scalerel}
\usepackage{tikz}
\usetikzlibrary{svg.path}

\definecolor{orcidlogocol}{HTML}{A6CE39}
\tikzset{
  orcidlogo/.pic={
    \fill[orcidlogocol] svg{M256,128c0,70.7-57.3,128-128,128C57.3,256,0,198.7,0,128C0,57.3,57.3,0,128,0C198.7,0,256,57.3,256,128z};
    \fill[white] svg{M86.3,186.2H70.9V79.1h15.4v48.4V186.2z}
                 svg{M108.9,79.1h41.6c39.6,0,57,28.3,57,53.6c0,27.5-21.5,53.6-56.8,53.6h-41.8V79.1z M124.3,172.4h24.5c34.9,0,42.9-26.5,42.9-39.7c0-21.5-13.7-39.7-43.7-39.7h-23.7V172.4z}
                 svg{M88.7,56.8c0,5.5-4.5,10.1-10.1,10.1c-5.6,0-10.1-4.6-10.1-10.1c0-5.6,4.5-10.1,10.1-10.1C84.2,46.7,88.7,51.3,88.7,56.8z};
  }
}

\newcommand\orcidicon[1]{\href{https://orcid.org/#1}{\mbox{\scalerel*{
\begin{tikzpicture}[yscale=-1,transform shape]
\pic{orcidlogo};
\end{tikzpicture}
}{|}}}}

%% file: tables/table.tex
\scalebox{0.6}
{
    \begin{tabular}{l cccc}
    \toprule 

    &

        \multicolumn{4}{c}{\textbf{Pressure}, $p$}
         \\
    
    \toprule 
    &

        \multicolumn{4}{c}{Training Set}
         \\
         Model &

         \textbf{PSNR $(\uparrow)$} &  
         \textbf{SSIM $(\uparrow)$} & 
         \textbf{R$^{2}$ $(\uparrow)$} &
         \textbf{R$^{2}_{r}$ $(\uparrow)$} 
         \\
        \midrule
         NF$^{*}$ &
         $ 18.6122 {\scriptstyle \pm 6.6895}$ &
         $ 0.8294 {\scriptstyle \pm 0.1568}$ &
         $ 0.2068 {\scriptstyle \pm 0.2438}$ & 
         $ 0.2815 {\scriptstyle \pm 0.2223}$ \\
         NF&
         $ 36.3476 {\scriptstyle \pm 0.6066}$ &
         $ 0.9916 {\scriptstyle \pm 0.0035}$ &
         $ 0.9925 {\scriptstyle \pm 0.0079}$ & 
         $ 0.9938 {\scriptstyle \pm 0.0021}$ \\
         NF-BC&
         $ 36.2597 {\scriptstyle \pm 0.9966}$ &
         $ 0.9896 {\scriptstyle \pm 0.0031}$ &
         $ 0.9838 {\scriptstyle \pm 0.0080}$ & 
         $ 0.9854 {\scriptstyle \pm 0.0071}$ \\
         NF-T&
         $ 37.7583 {\scriptstyle \pm 0.6121}$ &
         $ 0.9903 {\scriptstyle \pm 0.0031}$ &
         $ 0.9877 {\scriptstyle \pm 0.0066}$ & 
         $ 0.9893 {\scriptstyle \pm 0.0056}$ \\
         NF-BC-T&
         $ \textcolor{red}{40.8562 {\scriptstyle \pm 0.1928}}$ &
         $ \textcolor{red}{0.9950 {\scriptstyle \pm 0.0019}}$ &
         $ \textcolor{red}{0.9945 {\scriptstyle \pm 0.0061}}$ & 
         $ \textcolor{red}{0.9976 {\scriptstyle \pm 0.0053}}$ \\
        \midrule

         &

        \multicolumn{4}{c}{Test Set}
         \\
         Model &

         \textbf{PSNR $(\uparrow)$} &  
         \textbf{SSIM $(\uparrow)$} & 
         \textbf{R$^{2}$ $(\uparrow)$} &
         \textbf{R$^{2}_{r}$ $(\uparrow)$} 
         \\
        \midrule
         NF$^{*}$&
         $ 13.1425 {\scriptstyle \pm 3.5157}$ &
         $ 0.3866 {\scriptstyle \pm 0.2125}$ &
         $ 0 {\scriptstyle \pm 2.1789}$ & 
         $ 0 {\scriptstyle \pm 1.8253}$ \\
         NF&
         $ 32.3004 {\scriptstyle \pm 4.5651}$ &
         $ 0.9617 {\scriptstyle \pm 0.0230}$ &
         $ 0.9590 {\scriptstyle \pm 0.0322}$ & 
         $ 0.9617 {\scriptstyle \pm 0.0301}$ \\
         NF-BC&
         $ 31.1815 {\scriptstyle \pm 3.1450}$ &
         $ 0.9634 {\scriptstyle \pm 0.0189}$ &
         $ 0.9789 {\scriptstyle \pm 0.0178}$ & 
         $ 0.9807 {\scriptstyle \pm 0.0165}$ \\
         NF-T&
         $ 32.7285 {\scriptstyle \pm 2.7437}$ &
         $ 0.9654 {\scriptstyle \pm 0.0192}$ &
         $ 0.9885 {\scriptstyle \pm 0.0085}$ & 
         $ 0.9896 {\scriptstyle \pm 0.0078}$ \\
         NF-BC-T&
         $ \textcolor{red}{33.2971 {\scriptstyle \pm 3.8116}}$ &
         $ \textcolor{red}{0.9727 {\scriptstyle \pm 0.0149}}$ &
         $ \textcolor{red}{0.9890 {\scriptstyle \pm 0.0042}}$ & 
         $ \textcolor{red}{0.9952 {\scriptstyle \pm 0.0132}}$ \\
         \toprule 

    &

        \multicolumn{4}{c}{\textbf{Temperature}, $T$}
         \\
    
    \toprule 
    &

        \multicolumn{4}{c}{Training Set}
         \\
         Model &

         \textbf{PSNR $(\uparrow)$} &  
         \textbf{SSIM $(\uparrow)$} & 
         \textbf{R$^{2}$ $(\uparrow)$} &
         \textbf{R$^{2}_{r}$ $(\uparrow)$} 
         \\
        \midrule
         NF$^{*}$&
         $ 21.1829 {\scriptstyle \pm 5.7282}$ &
         $ 0.8127 {\scriptstyle \pm 0.1393}$ &
         $ 0.3318 {\scriptstyle \pm 0.2175}$ & 
         $ 0.7397 {\scriptstyle \pm 0.0787}$ \\
         NF&
         $ 35.9615 {\scriptstyle \pm 0.1014}$ &
         $ 0.9867 {\scriptstyle \pm 0.0043}$ &
         $ 0.984 {\scriptstyle \pm 0.0071}$ & 
         $ 0.9950 {\scriptstyle \pm 0.0015}$ \\
         NF-BC&
         $ 35.5010 {\scriptstyle \pm 0.5722}$ &
         $ 0.9852 {\scriptstyle \pm 0.0047}$ &
         $ 0.9778 {\scriptstyle \pm 0.0103}$ & 
         $ 0.9918 {\scriptstyle \pm 0.0031}$ \\
         NF-T&
         $ 39.8180 {\scriptstyle \pm 0.4639}$ &
         $ 0.9942 {\scriptstyle \pm 6.9606e-04}$ &
         $ 0.9960 {\scriptstyle \pm 6.9456e-04}$ & 
         $ 0.9983 {\scriptstyle \pm 1.9329e-04}$ \\
         NF-BC-T&
         $ \textcolor{red}{42.3375 {\scriptstyle \pm 0.1634}}$ &
         $ \textcolor{red}{0.9965 {\scriptstyle \pm 5.1562e-04}}$ &
         $ \textcolor{red}{0.9972 {\scriptstyle \pm 8.6986e-04}}$ & 
         $ \textcolor{red}{0.9989 {\scriptstyle \pm 2.6814e-04}}$ \\
        \midrule

         &

        \multicolumn{4}{c}{Test Set}
         \\
         Model &

         \textbf{PSNR $(\uparrow)$} &  
         \textbf{SSIM $(\uparrow)$} & 
         \textbf{R$^{2}$ $(\uparrow)$} &
         \textbf{R$^{2}_{r}$ $(\uparrow)$} 
         \\
        \midrule
         NF$^{*}$&
         $ 15.7772 {\scriptstyle \pm 1.7795}$ &
         $ 0.2497 {\scriptstyle \pm 0.2440}$ &
         $ 0.1032 {\scriptstyle \pm 0.4370}$ & 
         $ 0.5784 {\scriptstyle \pm 0.1970}$ \\
         NF&
         $ 31.9065 {\scriptstyle \pm 2.4295}$ &
         $ 0.9563 {\scriptstyle \pm 0.0202}$ &
         $ 0.9740 {\scriptstyle \pm 0.0163}$ & 
         $ 0.9890 {\scriptstyle \pm 0.0067}$ \\
         NF-BC&
         $ 31.2742 {\scriptstyle \pm 2.3244}$ &
         $ 0.9552 {\scriptstyle \pm 0.0200}$ &
         $ 0.9785 {\scriptstyle \pm 0.0142}$ & 
         $ 0.9921 {\scriptstyle \pm 0.0052}$ \\
         NF-T&
         $ 38.5099 {\scriptstyle \pm 0.8126}$ &
         $ 0.9924 {\scriptstyle \pm 0.0014}$ &
         $ 0.9960 {\scriptstyle \pm 8.5638e-04}$ & 
         $ 0.9981 {\scriptstyle \pm 3.8661e-04}$ \\
         NF-BC-T&
         $ \textcolor{red}{39.8183 {\scriptstyle \pm 1.3419}}$ &
         $ \textcolor{red}{0.9940 {\scriptstyle \pm 0.0016}}$ &
         $ \textcolor{red}{0.9969 {\scriptstyle \pm 0.0011}}$ & 
         $ \textcolor{red}{0.9986 {\scriptstyle \pm 4.8906e-04}}$ \\

         \toprule 

    &

        \multicolumn{4}{c}{\textbf{Horizontal Velocity}, $v_x$}
         \\
    
    \toprule 
    &

        \multicolumn{4}{c}{Training Set}
         \\
         Model &

         \textbf{PSNR $(\uparrow)$} &  
         \textbf{SSIM $(\uparrow)$} & 
         \textbf{R$^{2}$ $(\uparrow)$} &
         \textbf{R$^{2}_{r}$ $(\uparrow)$} 
         \\
        \midrule
         NF$^{*}$&
         $ 15.7796 {\scriptstyle \pm 3.4036}$ &
         $ 0.7483 {\scriptstyle \pm 0.0883}$ &
         $ 0 {\scriptstyle \pm 1.4052}$ & 
         $ 0.4376 {\scriptstyle \pm 0.3186}$ \\
         NF&
         $ 28.8057 {\scriptstyle \pm 0.3302}$ &
         $ 0.9777 {\scriptstyle \pm 0.0043}$ &
         $ 0.9744 {\scriptstyle \pm 0.0066}$ & 
         $ 0.9935 {\scriptstyle \pm 0.0012}$ \\
         NF-BC&
         $ 29.7342 {\scriptstyle \pm 0.6922}$ &
         $ 0.9777 {\scriptstyle \pm 0.0076}$ &
         $ 0.9748 {\scriptstyle \pm 0.0081}$ & 
         $ 0.9934 {\scriptstyle \pm 0.0018}$ \\
         NF-T&
         $ 28.4511 {\scriptstyle \pm 0.2283}$ &
         $ 0.9750 {\scriptstyle \pm 0.0065}$ &
         $ 0.9718 {\scriptstyle \pm 0.0065}$ & 
         $ 0.9926 {\scriptstyle \pm 0.0013}$ \\
         NF-BC-T&
         $ \textcolor{red}{32.3887 {\scriptstyle \pm 0.1468}}$ &
         $ \textcolor{red}{0.9864 {\scriptstyle \pm 0.0033}}$ &
         $ \textcolor{red}{0.9842 {\scriptstyle \pm 0.0061}}$ & 
         $ \textcolor{red}{0.9961 {\scriptstyle \pm 0.0013}}$ \\
        \midrule

         &

        \multicolumn{4}{c}{Test Set}
         \\
         Model &

         \textbf{PSNR $(\uparrow)$} &  
         \textbf{SSIM $(\uparrow)$} & 
         \textbf{R$^{2}$ $(\uparrow)$} &
         \textbf{R$^{2}_{r}$ $(\uparrow)$} 
         \\
        \midrule
         NF$^{*}$&
         $ 8.7470 {\scriptstyle \pm 2.9341}$ &
         $ 0.1849 {\scriptstyle \pm 0.2254}$ &
         $ 0 {\scriptstyle \pm 0.8020}$ & 
         $ 0.6255 {\scriptstyle \pm 0.2307}$ \\
         NF&
         $ 26.7855 {\scriptstyle \pm 1.4551}$ &
         $ 0.9458 {\scriptstyle \pm 0.0208}$ &
         $ 0.9659 {\scriptstyle \pm 0.0153}$ & 
         $ 0.9900 {\scriptstyle \pm 0.0045}$ \\
         NF-BC&
         $ 27.1669 {\scriptstyle \pm  1.6665}$&
         $ 0.9425 {\scriptstyle \pm 0.0239}$ &
         $ 0.9722 {\scriptstyle \pm 0.0149}$ & 
         $ 0.9928 {\scriptstyle \pm 0.0036}$ \\
         NF-T&
         $ 26.9675 {\scriptstyle \pm 1.0020}$ &
         $ 0.9470 {\scriptstyle \pm 0.0192}$ &
         $ 0.9756 {\scriptstyle \pm 0.0066}$ & 
         $ 0.9931 {\scriptstyle \pm 0.0017}$ \\
         NF-BC-T&
         $ \textcolor{red}{28.9846 {\scriptstyle \pm 1.8656}}$ &
         $ \textcolor{red}{0.9527 {\scriptstyle \pm 0.0210}}$ &
         $ \textcolor{red}{0.9800 {\scriptstyle \pm 0.0112}}$ & 
         $ \textcolor{red}{0.9948 {\scriptstyle \pm 0.0028}}$ \\
         \toprule 

    &

        \multicolumn{4}{c}{\textbf{Vertical Velocity}, $v_y$}
         \\
    
    \toprule 
    &

        \multicolumn{4}{c}{Training Set}
         \\
         Model &

         \textbf{PSNR $(\uparrow)$} &  
         \textbf{SSIM $(\uparrow)$} & 
         \textbf{R$^{2}$ $(\uparrow)$} &
         \textbf{R$^{2}_{r}$ $(\uparrow)$} 
         \\
        \midrule
         NF$^{*}$&
         $ 18.8795 {\scriptstyle \pm 3.7899}$ &
         $ 0.7303 {\scriptstyle \pm 0.0895}$ &
         $ 0 {\scriptstyle \pm 0.5140}$ & 
         $ 0.1299 {\scriptstyle \pm 0.3888}$ \\
         NF&
         $ 32.5914 {\scriptstyle \pm 1.9737}$ &
         $ 0.9495 {\scriptstyle \pm 0.0301}$ &
         $ 0.9450 {\scriptstyle \pm 0.0441}$ & 
         $ 0.9647 {\scriptstyle \pm 0.0274}$ \\
         NF-BC&
         $ 33.4592 {\scriptstyle \pm 2.2623}$ &
         $ 0.9531 {\scriptstyle \pm 0.0156}$ &
         $ 0.9590 {\scriptstyle \pm 0.0301}$ & 
         $ 0.9741 {\scriptstyle \pm 0.0194}$ \\
         NF-T&
         $ 31.6973 {\scriptstyle \pm 1.6549}$ &
         $ 0.9636 {\scriptstyle \pm 0.0096}$ &
         $ 0.9623 {\scriptstyle \pm 0.0227}$ & 
         $ 0.9769 {\scriptstyle \pm 0.0120}$ \\
         NF-BC-T&
         $ \textcolor{red}{34.7668 {\scriptstyle \pm 1.7010}}$ &
         $ \textcolor{red}{0.9709 {\scriptstyle \pm 0.0078}}$ &
         $ \textcolor{red}{0.9690 {\scriptstyle \pm 0.0304}}$ & 
         $ \textcolor{red}{0.9787 {\scriptstyle \pm 0.0156}}$ \\
        \midrule

         &

        \multicolumn{4}{c}{Test Set}
         \\
         Model &

         \textbf{PSNR $(\uparrow)$} &  
         \textbf{SSIM $(\uparrow)$} & 
         \textbf{R$^{2}$ $(\uparrow)$} &
         \textbf{R$^{2}_{r}$ $(\uparrow)$} 
         \\
        \midrule
         NF$^{*}$&
         $ 11.7643 {\scriptstyle \pm 2.3481}$ &
         $ 0.1756 {\scriptstyle \pm 0.1929}$ &
         $ 0.3230 {\scriptstyle \pm 0.2976}$ & 
         $ 0.5031 {\scriptstyle \pm 0.1994}$ \\
         NF&
         $ 24.9929 {\scriptstyle \pm 3.5585}$ &
         $ 0.8076 {\scriptstyle \pm 0.1109}$ &
         $ 0.9529 {\scriptstyle \pm 0.0315}$ & 
         $ 0.9754 {\scriptstyle \pm 0.0155}$ \\
         NF-BC&
         $ 25.9602 {\scriptstyle \pm 3.3207}$ &
         $ 0.8378 {\scriptstyle \pm 0.0877}$ &
         $ 0.9608 {\scriptstyle \pm 0.0266}$ & 
         $ 0.9817 {\scriptstyle \pm 0.0118}$ \\
         NF-T&
         $ 25.9108 {\scriptstyle \pm 2.7349}$ &
         $ \textcolor{red}{0.8543 {\scriptstyle \pm 0.0752}}$ &
         $ 0.9597 {\scriptstyle \pm 0.0254}$ & 
         $ 0.9810 {\scriptstyle \pm 0.0097}$ \\
         NF-BC-T&
         $ \textcolor{red}{26.0760 {\scriptstyle \pm 3.4673}}$ &
         $ 0.8462 {\scriptstyle \pm 0.0896}$ &
         $ \textcolor{red}{0.9610 {\scriptstyle \pm 0.0272}}$ & 
         $ \textcolor{red}{0.9887 {\scriptstyle \pm 0.0124}}$ \\
         
         \bottomrule
         
    \end{tabular}
}